\documentclass[12pt]{article} 
\usepackage{ucs} 
\usepackage[utf8x]{inputenc} 
\usepackage[T1]{fontenc} 
\usepackage{amsmath,amssymb} 
\usepackage{graphicx} 
\usepackage[unicode]{hyperref} 
\hypersetup{bookmarksopen=true, bookmarksnumbered=true, pdfstartview={FitH}, pdfborder={0 0 0}, colorlinks=true, linkcolor=red, linktocpage=true, citecolor=blue, urlcolor=cyan, unicode=true, pdftitle={Free Energy of Dn Quiver Chern-Simons Theories}, pdfauthor={PMCrichigno, CPHerzog, DJain}, pdfsubject={Free Energy Calculation}, pdfkeywords={Chern-Simons Theories, }{Matrix Models, }{Localization, }{Quivers}}
\PrerenderUnicode{₅}
\usepackage{geometry} 
\usepackage{fancyhdr} 
\usepackage[nottoc,notlof,notlot]{tocbibind} 
\usepackage[titles]{tocloft} 

\usepackage{parskip} 
\usepackage{array} 
\newcolumntype{V}{>{\centering\arraybackslash} m{.2\linewidth} }
\newcommand\references[1] {\begin{thebibliography}{[99]} #1 \end{thebibliography}}
\numberwithin{equation}{section} 
\newcommand\cen[1] {\begin{center}#1\end{center}}
\newcommand\equ[1] {\begin{equation}#1\end{equation}}
\newcommand\equn[1] {\begin{equation*}#1\end{equation*}}
\newcommand\eqs[1] {\begin{align}#1\end{align}}
\newcommand\eqsn[1] {\begin{align*}#1\end{align*}}
\newcommand\pmat[1] {\begin{pmatrix}#1\end{pmatrix}}
\renewcommand\( {\left(}
\renewcommand\) {\right)}

\DeclareUnicodeCharacter{"0393}{\Gamma}
\DeclareUnicodeCharacter{"0394}{\Delta}
\DeclareUnicodeCharacter{"0398}{\Theta}
\DeclareUnicodeCharacter{"039B}{\Lambda}
\DeclareUnicodeCharacter{"039E}{\Xi}
\DeclareUnicodeCharacter{"03A0}{\Pi}
\DeclareUnicodeCharacter{"03A3}{\Sigma}
\DeclareUnicodeCharacter{"03A5}{\Upsilon}
\DeclareUnicodeCharacter{"03A6}{\Phi}
\DeclareUnicodeCharacter{"03A8}{\Psi}
\DeclareUnicodeCharacter{"03A9}{\Omega}
\DeclareUnicodeCharacter{"03B1}{\alpha}
\DeclareUnicodeCharacter{"03B2}{\beta}
\DeclareUnicodeCharacter{"03B3}{\gamma}
\DeclareUnicodeCharacter{"03B4}{\delta}
\DeclareUnicodeCharacter{"03B5}{\epsilon}
\DeclareUnicodeCharacter{"03B6}{\zeta}
\DeclareUnicodeCharacter{"03B7}{\eta}
\DeclareUnicodeCharacter{"03B8}{\theta}
\DeclareUnicodeCharacter{"03D1}{\vartheta}
\DeclareUnicodeCharacter{"03B9}{\iota}
\DeclareUnicodeCharacter{"03BA}{\kappa}
\DeclareUnicodeCharacter{"03BB}{\lambda}
\DeclareUnicodeCharacter{"03BC}{\mu}
\DeclareUnicodeCharacter{"03BD}{\nu}
\DeclareUnicodeCharacter{"03BE}{\xi}
\DeclareUnicodeCharacter{"03C0}{\pi}
\DeclareUnicodeCharacter{"03C1}{\rho}
\DeclareUnicodeCharacter{"03C3}{\sigma} 
\DeclareUnicodeCharacter{"03C4}{\tau}
\DeclareUnicodeCharacter{"03C5}{\upsilon}
\DeclareUnicodeCharacter{"03C6}{\phi}
\DeclareUnicodeCharacter{"03D5}{\varphi}
\DeclareUnicodeCharacter{"03C7}{\chi}
\DeclareUnicodeCharacter{"03C8}{\psi}
\DeclareUnicodeCharacter{"03C9}{\omega}
\DeclareUnicodeCharacter{"21D0}{\Leftarrow}
\DeclareUnicodeCharacter{"0212B}{\AA}
\DeclareUnicodeCharacter{"266A}{\eighthnote}
\DeclareUnicodeCharacter{"266B}{\twonotes}

\newcommand\arXivid[1] {\href{http://arxiv.org/abs/#1}{\tt arXiv:#1}} 
\newcommand\cmp[3] {{\it Commun.\ Math.\ Phys.\ } \href{http://inspirehep.net/search?ln=en&ln=en&p=find+j+"Commun.Math.Phys.,#1,#3"&of=hb&action_search=Search&sf=&so=d&rm=&rg=25&sc=0}{{\bf #1} (#2) #3}} 

\newcommand\jhep[3]{{\it JHEP\ } \href{http://inspirehep.net/search?ln=en&ln=en&p=find+j+"JHEP,#1,#3"&of=hb&action_search=Search&sf=&so=d&rm=&rg=25&sc=0}{{\bf #1} (#2) #3}}
\newcommand\npb[3] {{\it Nucl.\ Phys.\ }{\bf B #1} (#2) #3}
\newcommand\pr[4] {{\it Phys.\ Rev.\ }{\bf #1 #2} (#3) #4} 
\newcommand\pl[4] {{\it Phys.\ Lett.\ }{\bf #1 #2} (#3) #4}

\begin{document}
\pagenumbering{alph}
\title{\Huge Free Energy of $\widehat D_n$ Quiver Chern-Simons Theories}
\author{P. Marcos Crichigno\footnote{\href{mailto:crichigno@insti.physics.sunysb.edu}{crichigno@insti.physics.sunysb.edu}}\,, Christopher P.  Herzog\footnote{\href{mailto:Christopher.Herzog@stonybrook.edu}{Christopher.Herzog@stonybrook.edu}}\,,  and Dharmesh Jain\footnote{\href{mailto:djain@ic.sunysb.edu}{djain@ic.sunysb.edu}}\bigskip\\ \emph{C. N. Yang Institute for Theoretical Physics}\\ \emph{State University of New York, Stony Brook, NY 11790-3840}}
\date{} 
\maketitle

\thispagestyle{fancy}
\rhead{YITP-SB-12-$41$} 
\lhead{\today}
\begin{abstract}
\normalsize We apply the matrix model of Kapustin, Willett and Yaakov to compute the free energy of $\mathcal N=3$ Chern-Simons matter theories with $\widehat D_n$ quivers in the large $N$ limit. 
We conjecture a general expression for the free energy that is explicitly invariant under Seiberg duality and show that it can be interpreted as a sum over certain graphs known as signed graphs. Through the AdS/CFT correspondence, this  leads to a prediction for the volume of certain tri-Sasaki Einstein manifolds. We also study the unfolding procedure, which relates these $\widehat D_n$ quivers to $\widehat A_{2n-5}$ quivers. Furthermore, we consider the addition of massive fundamental flavor fields, verifying that integrating these out decreases the free energy  in accordance with the F-theorem.  
\end{abstract}

\newpage
\pagenumbering{Roman}
\cfoot{\thepage}\rhead{}\lhead{}
\tableofcontents

\newpage
\pagenumbering{arabic}
\section{Introduction}
Supersymmetric localization \cite{Witten:1988ze,Pestun:2007rz} is a powerful method that makes exact computations in superconformal field theories (SCFTs) possible. This procedure reduces the infinite dimensional path integral to a finite dimensional integral, typically over  the Coulomb branch. It has recently been used to obtain interesting results for field theories in various number of dimensions \cite{Pestun:2007rz,Kapustin:2009kz,Jafferis:2010un,Hama:2010av,Kallen:2012cs,Jafferis:2012iv}. In particular, Kapustin, Willett and Yaakov \cite{Kapustin:2009kz}  applied localization in three dimensions to calculate the exact partition function on $S^3$ for theories with $\mathcal N \geq 2$ supersymmetry.  One of the outcomes of these calculations has been the proposal \cite{Jafferis:2011zi}  that the free energy, defined by
\equ{F=-\log|Z_{S^3}|\,,
\label{F log Z}
} 
decreases along renormalization group (RG) flows, providing a good measure of the degrees of freedom in the field theory. On the other hand, many  three-dimensional SCFTs can be realized as effective theories of coincident $N$ M2-branes. Thus, localization is a tool that can be used to test predictions of the AdS/CFT correspondence \cite{Maldacena:1997re,Gubser:1998bc,Witten:1998qj}. One of the first and most remarkable results \cite{Drukker:2010nc} was the evaluation of the free energy for $U(N)_k \times U(N)_{-k}$ Chern-Simons (CS) theory \cite{Aharony:2008ug}, matching the famous $N^{3/2}$ scaling of gravitational free energy predicted in \cite{Klebanov:1996un}.

A larger class of quiver Chern-Simons theories  with $U(N)_{k_1}\times U(N)_{k_2}\times ... \times U(N)_{k_n}$ gauge groups, coupled to bifundamental matter forming a necklace-type quiver  were considered in \cite{Jafferis:2011zi,Herzog:2010hf, Martelli:2011qj, Cheon:2011vi}. It is believed that the M-theory description of these theories \cite{Jafferis:2008qz} arises as the near-horizon limit of a stack of $N$ M2-branes placed at the tip of a Calabi-Yau cone with a tri-Sasaki Einstein base $Y$. In the large $N$ limit, the gravitational free energy is given by \cite{Drukker:2010nc,Herzog:2010hf} 
\equ{F=N^{3/2}\sqrt{\frac{2 \pi^6}{27\,\text{Vol}(Y)}}+o(N^{3/2})\,,
\label{F Vol}
}
where $\text{Vol}(Y)$ is the volume of the compact manifold $Y$ whose geometry depends on the quiver data, in particular the CS levels. By evaluating the free energy (\ref{F log Z}) for the necklace quivers and matching it with the gravitational energy, an expression for $\text{Vol}(Y)$ as a function of the CS levels  was conjectured in \cite{Herzog:2010hf}. This was corroborated in \cite{Gulotta:2011si} by comparison with the explicit calculations  of the volumes of toric Sasaki-Einstein manifolds \cite{Yee:2006ba} (see also \cite{Assel:2012cj} for a calculation in type-IIB supergravity).
  
These necklace quivers are actually an example of a more general class of quiver theories which have a nice large $N$ limit, $i.e.$, long-range forces between eigenvalues in the matrix model cancel \cite{Gulotta:2011vp}.  In fact, quiver theories for which this happens are in one-to-one correspondence with the extended ADE Dynkin diagrams with necklace quivers corresponding to the $\widehat A$-class. 

In this paper we focus on theories with $\widehat D_n$ quivers. The relevant tri-Sasaki Einstein manifold $Y$ is the base of the hyperk\"{a}hler cone $\mathbb H^{4n-8}/// U(1)^{n-1} \times SU(2)^{n-3}$. By assuming a particular ordering of CS levels, we solve the matrix models for various values of $n$ and propose an expression for $\text{Vol}(Y)$ for arbitrary $n$. This expression is related to the area of a certain polygon as in the case of $\widehat A$-quivers. Then, we propose a general expression given by a  rational function, which is valid for any ordering of the CS levels and is invariant under Seiberg duality.  The numerator for such an expression was given in \cite{Gulotta:2011vp}.  Here we give the denominator as well, leading to the conjecture
\equ{\frac{\text{Vol}(Y)}{\text{Vol}(S^7)} = \frac{\sum_{\mathcal{R}_+} \det (\alpha_1...\alpha_n)^2 \prod _{a=1}^n |\alpha_a \cdot p |}{8(n-2)\(\sum_{a=1}^n |p_n|\) \prod_{a=1}^{n}\left[ \sum_{b=1}^{n}\(|p_a-p_b|+|p_a+p_b|\)-4|p_a| \right] }\,,
\label{Vol D intro}
}
where $\mathcal R_+$ is an $n$-subset of positive roots $\alpha_a$ of $D_n$ and the CS levels are $k_a=\alpha_a \cdot p$. We show that the numerator of this expression can be expressed as a sum over certain graphs known as \textit{signed graphs}. Using a generalized matrix-tree formula, we show that (\ref{Vol D intro}) reduces to the polygon formula for a particular ordering of the CS levels. Although we do not discuss exceptional quivers in detail,  we give the free energy for $\widehat E_6, \widehat E_7, \widehat E_8$ in Appendix~\ref{Exceptional Quivers} for completeness. 

This paper is organized as follows: In Section~\ref{ADE Matrix Models} we review the matrix model, the large $N$ limit and the ADE classification. In Section~\ref{Solving the Matrix Models} we explicitly solve the $\widehat D_n$ matrix models for $n=5\,, 6$ and conjecture an expression for arbitrary $n$. In Section~\ref{Free Energy for Dn} we propose the general formula and the relation to signed graphs. In Section~\ref{Flavored $D_n$ quivers and the F-theorem} we add fundamental flavor fields and check the F-theorem. In Section~\ref{Unfolding section} we apply the procedure known as unfolding, which relates the free energy for $\widehat D$-quivers to that of $\widehat A$-quivers for a particular choice of CS levels. We conclude with a summary and discussion of open problems.

\section{\texorpdfstring{$\widehat{ADE}$ Matrix Models}{ADE Matrix Models}}\label{ADE Matrix Models}
We will consider quiver Chern-Simons gauge theories involving products of unitary groups only, $i.e.$, $G=\otimes_aU(n_a N)$, coupled to bifundamental chiral superfields $(A_a,\, B_a)$. According to \cite{Kapustin:2009kz}, the partition function of these theories on $S^3$ is localized on configurations where the auxiliary scalar fields $\sigma_a$ in the $\mathcal N=2$ vector multiplets are  constant $N \times N$ matrices. Thus, evaluating the free energy amounts to solving a matrix model. 

\subsection*{Matrix Model}
We denote the eigenvalues of $\sigma_a$ in each vector multiplet by $\lambda_{a,i}$, $i=1,...,N_a$. The partition function is then given by \cite{Kapustin:2009kz}
\equ{Z = \int \( \prod_{a, i} d \lambda_{a,i} \) L_v(\{\lambda_{a,i}\}) L_m(\{\lambda_{a,i} \}) = \int \( \prod_{a, i} d \lambda_{a,i} \)  \exp\left[-F (\{ \lambda_{a,i} \}) \right],
}
where the contribution from vector multiplets is
\equn{L_v = \prod_{a=1}^d \frac{1}{N_a!} \( \prod_{i > j} 2 \sinh[ \pi (\lambda_{a,i} - \lambda_{a,j})] \)^2 \exp \(i \pi \sum_{a,j} k_a \lambda_{a,j}^2 \),
}
and from matter multiplets is
\equn{L_m = \prod_{(a,b) \in E} \prod_{i,j} \frac{1}{2 \cosh [ \pi (\lambda_{a,i} - \lambda_{b,j})]} \prod_c \(\prod_i \frac{1}{2 \cosh[\pi \lambda_{c,i}]}\)^{n^f_c}.
} 
The first product in $L_m$ is due to bifundamental fields while the second one is due to fundamental flavor fields, where $n^f_c$ is the number of pairs of flavor fields at the node labeled by the index $c$.

\subsection*{\texorpdfstring{Large $N$ Limit and $\widehat{ADE}$ Classification}{Large N Limit and ADE Classification}}
Following \cite{Herzog:2010hf,Gulotta:2011vp}, we assume that the eigenvalue distribution becomes dense in the large $N$ limit, $i.e.$, $\lambda_{a,i} \rightarrow \lambda_a(x)$ with a certain density $\rho(x)$. In this limit  the free energy becomes a 1-dimensional integral which we evaluate by saddle point approximation. We also assume that the eigenvalue distribution for a node with $N_a=n_a N$ is given by a collection of $n_a$ curves in the complex plane labeled by $\lambda_{a,I}(x)$ with $I=1, ...,n_a$ and write the ansatz 
\equ{\lambda_{a,I}(x)=N^\alpha x +i\, y_{a,I}(x)\,.
\label{Ansatz lambda}
}
The density $\rho(x)$ is assumed to be normalized, $i.e.$,
\equ{\int dx \rho(x)=1\,,
\label{Normalization rho}
}
which will be imposed through a Lagrange multiplier $\mu$. As explained in \cite{Gulotta:2011vp}, the leading order in $N$ in the saddle point equation is proportional to the combination $2 n_a -\sum_{b| (a,b)\in E} n_b$. The requirement that this term vanishes is equivalent to the quiver being in correspondence with simply laced extended Dynkin diagrams, leading to the ADE classification. To next order in $N$, the saddle point equation contains a tree-level contribution and a 1-loop contribution. Assuming that $\sum_a n_a k_a=0$, the requirement that these two contributions are balanced leads to $\alpha=1/2$, which is ultimately responsible for the $N^{3/2}$ scaling of the free energy\footnote{Alternatively, one can assume that $\sum n_a k_a \neq 0$, and choose $\alpha=1/3$, which leads to a massive IIA supergravity dual \cite{Jafferis:2011zi}. We will not consider this case here.}. Finally, the Lagrangian to be extremized reads
\eqs{F = N^{3/2} &\int \rho(x) \Biggl[ \pi n_F |x|+ 2 \pi x \sum_a \sum_{I=1}^{n_a} k_a y_{a,I}(x)
+ \frac{ \rho(x)}{4\pi} \Biggl(\sum_{a=1}^d \sum_{I=1}^{n_a} \sum_{J=1}^{n_a} \arg \(e^{2 \pi i (y_{a,I} - y_{a,J} - 1/2)} \)^2 \nonumber\\
&-\sum_{(a,b) \in E} \sum_{I=1}^{n_a} \sum_{J=1}^{n_b} \arg \( e^{2 \pi i (y_{a,I} - y_{b,J})} \)^2\Biggr)
\Biggr] dx - 2 \pi \mu N^{3/2} \( \int  \rho(x) \, dx- 1 \),
\label{F}
}
where  $n_F \equiv \sum_a n_a^f n_a$. Evaluating the free energy on-shell gives
\equ{F=\frac{4 \pi N^{3/2}}{3} \mu\,,
\label{Fproptomu}
}
which can be understood as a virial theorem \cite{Gulotta:2011si}. Thus, the free energy is determined by $\mu$, which in turn is determined as a function of the CS levels from the normalization condition (\ref{Normalization rho}). Note that from (\ref{F Vol}) and (\ref{Fproptomu}), it follows that
\equ{\frac{\text{Vol}(Y)}{\text{Vol}(S^7)}=\frac{1}{8\,\mu^2}\,.
\label{relation vol and mu}
}

\begin{figure}[h]
\centering
\includegraphics[width=3in]{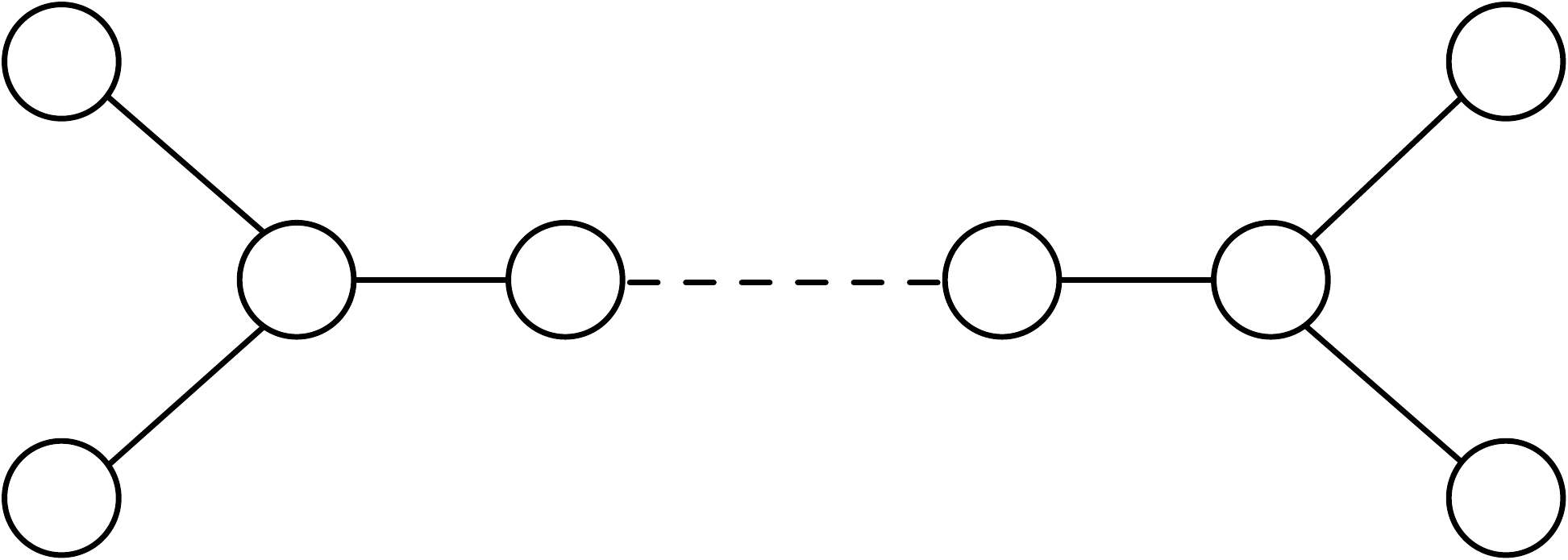}
\put (1,4) {$k_3$} \put (1,65) {$k_4$}\put (-228,65) {$k_1$} \put (-228,4) {$k_2$} \put (-180,50) {$k_5$} \put (-142,50) {$k_6$} \put (-115,50) {$\cdots$} \put (-58,50) {$k_{n+1}$}
\caption{$\widehat D_n$ quiver diagram. Each node `$a$'  corresponds to a $U(n_a N)$ gauge group with CS level $k_a$, where $n_a$ is the node's comark  and we assume that $\sum_a n_a k_a=0$.}
\label{Dn}
\end{figure}
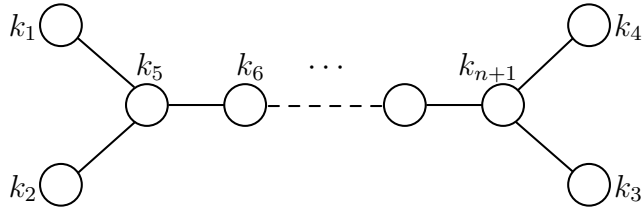

As mentioned earlier, theories with $\widehat A_{m-1}$ quiver diagrams have been extensively studied. Here, we wish to study theories with $\widehat D_n$ quivers as the one shown in Fig.~\ref{Dn}. For now we will set $n_a^f=0$, but we will reintroduce flavors in Section \ref{Flavored $D_n$ quivers and the F-theorem}.

It is convenient to relate the CS level $k_{(a)}$ at each node  to a root $\alpha_a$, by  introducing a vector $p$ and writing $k_{(a)}=\alpha_a \cdot p\,$. This way, the condition $\sum_a n_a k_a=0$ is satisfied automatically. Choosing a basis for the roots of $\widehat D_n$ (see Appendix~\ref{A and D Appendix} for conventions), we have
\begin{gather}
k_1=-(p_1+p_2)\,, \quad k_2 = p_1-p_2\,,\quad k_3=p_{n-1}-p_n\,, \quad k_4 = p_{n-1}+p_n\,,\nonumber\\
k_{i}=p_{i-3}-p_{i-2}\,; \qquad  i=5,...,n+1\,.
\label{k2p}
\end{gather}

 In the next two sections we will solve the matrix models for various $\widehat D_n$ quivers and conjecture a general volume formula for arbitrary $n$.

\section{Solving the Matrix Models}\label{Solving the Matrix Models}
Here we describe the saddle point evaluation of the free energy (\ref{F}). We show in detail the solution for $n=5$, state the result for $n=6$, and propose a general expression that we have checked for $n=7,...,10$. Finally, we will relate this expression to the area of a certain polygon.
 
\subsection{Explicit Solutions} 
\begin{figure}[h]
\centering
\begin{tabular}{cc}
\includegraphics[width=0.48\textwidth]{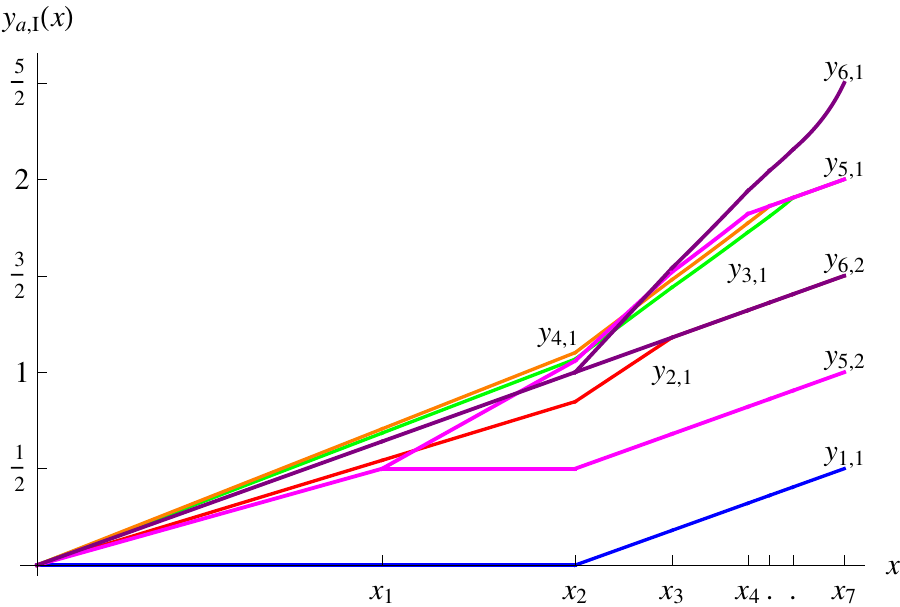} & \includegraphics[width=0.48\textwidth]{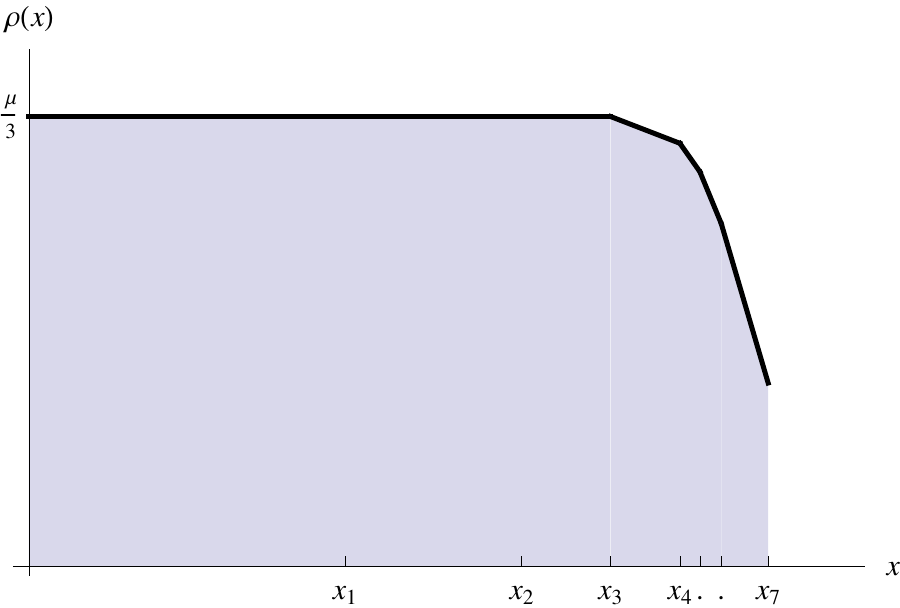}
\end{tabular}
\caption{The eigenvalue distribution $y_{a,I}(x)$ (left) for all nodes and density $\rho(x)$ (right) for the $\widehat D_5$ quiver with CS levels: $ (k_2,k_3,k_4,k_5,k_6) = (2,2,3,4,4)$.}
\label{Ys and rho D5}
\end{figure}

Extremizing (\ref{F}) (with respect to $y_{a,I}$ and $\rho$) requires an assumption on the branch of the arg functions. We will always take the principle value and therefore we assume that
\eqs{| y_{a,I} - y_{a,J}| < 1\,; \qquad | y_{a,I} - y_{b,J}| < \frac{1}{2}\,, \quad \text{if} \,\, (a,b) \in E \,.
\label{inequalities}
}
Based on numerical results \cite{Herzog:2010hf, Gulotta:2011vp}, we assume that the $n_a$ curves for a given node initially coincide, $i.e.$, $|y_{a,I}-y_{a,J}|=0$. Extremizing $F$ under these assumptions, one finds that the solution is consistent only in a bounded region away from the origin. This is because as $|x|$ increases, the differences $|y_{a,I} - y_{b,J}|$ monotonically increase (or decrease), saturating  one (or more) of the inequalities assumed in (\ref{inequalities}) at some point. The relation among the CS levels determines the sequence in which these inequalities saturate. This saturation will be maintained beyond this point, requiring the eigenvalue distribution involved either to bifurcate or develop a kink. As an example, consider the first plot in Fig.~\ref{Ys and rho D5} where we show the eigenvalue distributions for the $\widehat D_5$ quiver\footnote{We have used the freedom to add an arbitrary function to the $y_{a,I}$ to set $y_{1,1}(x)=0$ in the first region and we solve explicitly only for $x\geq0$ since the eigenvalue distributions and density are even functions of $x$.}. It consists of seven regions determined by saturation of different inequalities. At the end of first region ($x=x_1$), one can see that $y_{1,1}-y_{5,2}=-1/2$ forcing $y_{5,1}$ and $y_{5,2}$ to bifurcate. 

After a saturation occurs, the total number of independent variables is reduced by one. Thus, at this point, we remove one variable from the Lagrangian, revise the inequalities and solve the equations of motion again until a new saturation is encountered. This process is iterated until all $y_a$'s are related, determining a maximum of $(\sum_a n_a-1)$ regions or until the eigenvalue distribution terminates, $i.e.$, $\rho(x)=0$.  Once the eigenvalue density $\rho(x)$ is determined in all regions, the value of $\mu$ (and therefore $F$) is found from the normalization condition (\ref{Normalization rho}).

The solution to the $\widehat D_4$ quiver  consists of five regions and was solved in \cite{Gulotta:2011vp}. Assuming that $p_1\geq p_2\geq p_3 \geq p_4 \geq0$, it was found that
\eqs{\frac{1}{\mu^2}=-\frac{1}{4 p_1}+\frac{2 p_1+3 p_2-p_3}{(p_1+p_2)^2}-\frac{1}{2 (p_1+p_2+p_3-p_4)}-\frac{1}{2 (p_1+p_2+p_3+p_4)}\,.
\label{Sol D4}
}

We now discuss the solution to the $\widehat D_5$ quiver, consisting of seven regions. We assume that $k_6 \geq k_5 \geq k_4 \geq k_3 \geq k_2 \geq 0$  with $k_1=-(k_2+k_3+k_4+2k_5+2k_6)$ implying $p_1\geq p_2 \geq p_3 \geq p_4 \geq p_5 \geq 0$. The solution is summarized in Table \ref{Table D5} and Fig. \ref{Ys and rho D5} shows the eigenvalue distributions and density  (further details are given in Appendix \ref{D5App}).
\begin{table}[t]
\centering
\begin{tabular}{|c|c|c|c|} \hline
Region & $x_{i} $& $\delta y(x=x_{i})$ & $\rho_i(x)$\\ \hline \hline
1& $\frac{ \mu }{3 (k_2+k_3+k_4+2 k_5+2 k_6)}$ & $y_{1,1}-y_{5,2}=-\frac{1}{2}$ & $ \frac{1}{3} \mu$ \\ \hline
2& $\frac{4 \mu }{6 k_2+9 k_3+9 k_4+12 k_5+18 k_6}$ &  $y_{5,2}-y_{6,2}=-\frac{1}{2}$ & $\frac{1}{3} \mu$ \\ \hline
3& $ \frac{2 \mu }{3 (2 k_2+k_3+k_4+2 k_5+2 k_6)} $& $y_{2,1}-y_{6,2}=0$ & $ \frac{1}{3} \mu$ \\ \hline
4& $\frac{2 \mu }{2 k_2+3 (k_3+k_4+2 k_5+2 k_6)}$ & $y_{5,1}-y_{6,2}=\frac{1}{2}$ & $\frac{1}{2}\mu + \frac{1}{4} x (k_1 - k_2)$ \\ \hline
5&$\frac{2 \mu }{2 k_2+3 k_3+5 k_4+4 k_5+6 k_6}$ &  $y_{4,1}-y_{6,2}=\frac{1}{2}$ & $ \mu + x k_1$ \\ \hline
6& $\frac{2 \mu }{2 k_2+5 k_3+3 k_4+4 k_5+6 k_6}$ &  $y_{3,1}-y_{6,2}=\frac{1}{2}$ & $\frac{3}{2}\mu + \frac{1}{4} x (6 k_1 - k_3 - 3 k_4 -2 k_6)$ \\ \hline
7&  $\frac{2 \mu }{2 k_2+3 k_3+3 k_4+4 k_5+6 k_6}$ &  $y_{6,1}-y_{6,2}=1$ & $2 \mu + x (2 k_1 - k_3 - k_4 - k_6)$ \\ \hline
\end{tabular}
\caption{Key characteristics of the seven regions of the $\widehat D_5$ matrix model: their boundaries, the saturated inequalities and the eigenvalue densities, assuming  $k_6 \geq k_5 \geq k_4 \geq k_3 \geq k_2 \geq 0$.}
\label{Table D5}
\end{table}
From the information given in Table~\ref{Table D5}  and (\ref{Normalization rho}), we find
\eqs{\nonumber \frac{1}{\mu^2} =& -\frac{1}{2k_2+5k_3+3k_4+4k_5+6k_6} - \frac{1}{2 k_2 +3k_3+5k_4+4 k_5+6 k_6}\\ \nonumber
&+\frac{4(k_3+k_4+3k_6-2k_1)}{(2k_2+3k_3+3k_4+4k_5+6k_6)^2}\\ \nonumber
&-\frac{1}{9(2k_2+k_3+k_4+2k_5+2k_6)}-\frac{1}{2k_2 +3k_3 +3k_4 +6k_5 +6k_6} \,, }
which, using the relations in (\ref{k2p}) gives
\eqs{ \frac{1}{\mu^2} =& -\frac{1}{18 p_1} -\frac{1}{2( p_1 + 2 p_2)} +\frac{(2 p_1 + 2 p_2 + 3 p_3 - p_4)}{(p_1 + p_2 + p_3)^2}\nonumber\\
&-\frac{1}{2 (p_1 + p_2 + p_3 + p_4 - p_5)} - \frac{1}{2 (p_1 + p_2 + p_3 + p_4 + p_5)}\,.
\label{Sol D5}
}

Similarly, solving the $\widehat D_6$ matrix model as described above leads to a total of nine regions and integrating the eigenvalue density gives
\eqs{\frac{1}{\mu^2}=& -\frac{1}{48 p_1}-\frac{1}{6 (p_1+3 p_2)}-\frac{1}{2 (p_1+p_2+2 p_3)} + \frac{2 (p_1+p_2+p_3)+3 p_4-p_5}{(p_1+p_2+p_3+p_4)^2}\nonumber\\
&-\frac{1}{2 (p_1+p_2+p_3+p_4+p_5-p_6)}-\frac{1}{2 (p_1+p_2+p_3+p_4+p_5+p_6)}\,,
\label{Sol D6}
}
for $p_1\geq p_2 \geq ... \geq p_6 \geq 0$.

\subsection{General Solution and Polygon Area}\label{General solution and polygon area}
By comparing (\ref{Sol D4}), (\ref{Sol D5}) and (\ref{Sol D6}), we propose that the free energy for $\widehat D_n$ quivers  is determined by:
\eqs{\nonumber \frac{1}{\mu^2} =&\,\frac{1}{2}\sum_{a=1}^{n-3}\frac{c_a}{\sum_{b=1}^{a-1}p_b +(n-a-1)p_{a}}  + \frac{2\sum_{b=1}^{n-3} p_b +3 p_{n-2}-p_{n-1}}{\(\sum_{b=1}^{n-2} p_b\)^2}\\
&-\frac{1}{2}\(\frac{1}{\sum_{b=1}^{n-1}p_b -p_n}+\frac{1}{\sum_{b=1}^n p_b} \),
\label{Partial fractions Dn}
}
with $c_a\equiv  \frac{-2}{(n-a-1)(n-a-2)}$ and $p_1 \geq p_2 \geq... \geq p_n>0$. We have verified that this is correct for the $\widehat D_7, ..., \widehat D_{10}$ matrix models. 

For $\widehat A$-quivers, it was shown in \cite{Gulotta:2011si} that $\text{Vol}(Y)$ can be interpreted as the area of a certain polygon. By rewriting (\ref{Partial fractions Dn}) in a more suggestive form, we will show that there is a certain polygon (or rather a cone) whose area is related to $\text{Vol}(Y)$ for $\widehat D$-quivers as well. This construction will be particularly useful in Sections \ref{Flavored $D_n$ quivers and the F-theorem} and \ref{Unfolding section}. We start by observing that the denominators appearing in (\ref{Partial fractions Dn}) can be written as
\begin{gather}
\bar\sigma_{a} = \sum_{b=1}^{n}\(|p_a-p_b|+|p_a+p_b|\)-4\, |p_a|\,; \quad a=1,...,n\,,\nonumber\\
\bar \sigma_0 = 2(n-2)\,, \qquad \bar \sigma_{n+1}=2 \sum_{b=1}^n |p_b|\,.
\label{def sigmas}
\end{gather}
The first step in rewriting (\ref{Partial fractions Dn})  is to combine consecutive terms to get
\equ{\frac{\text{Vol}(Y)}{\text{Vol}(S^7)}=\frac{1}{2} \(\frac{1}{\bar \sigma_0\, \bar \sigma_1} + \sum_{a=1}^{n-1} \frac{p_{a}-p_{a+1}}{\bar \sigma_a\, \bar \sigma_{a+1}} +\frac{p_n}{ \bar \sigma_n\, \bar \sigma_{n+1}}\),
}
where we have used the relation (\ref{relation vol and mu}). The next step is to introduce the vectors $\beta_a= (1, p_a)$ together with $\beta_0=(0,1)$ and $\beta_{n+1}=(1,0)$. Defining the wedge product $(a,b) \wedge (c,d) = (a d- b c)$, we can write all the $\bar \sigma_a$'s in (\ref{def sigmas}) in terms of $\gamma_{a,b}\equiv \left|  \beta_a \wedge \beta_b \right|$ as follows
\equ{\bar \sigma_a  = \sum_{b=1}^n \( \gamma_{a,b}+ \gamma_{a,-b} \) - 4 \gamma_{a,n+1}\,; \quad a=0,...,n+1\,,
\label{all sigmas}
}
where we have also defined $\beta_{-a} \equiv (1,-p_a)$. This finally leads  to
\equ{\frac{\text{Vol}(Y)}{\text{Vol}(S^7)}=\frac{1}{2} \sum_{a=0}^{n} \frac{\gamma_{a,a+1}}{\bar \sigma_a\, \bar \sigma_{a+1}}\,.
\label{Dn sigmas}
}

\begin{figure}[h]
\centering
\includegraphics[width=1.5in]{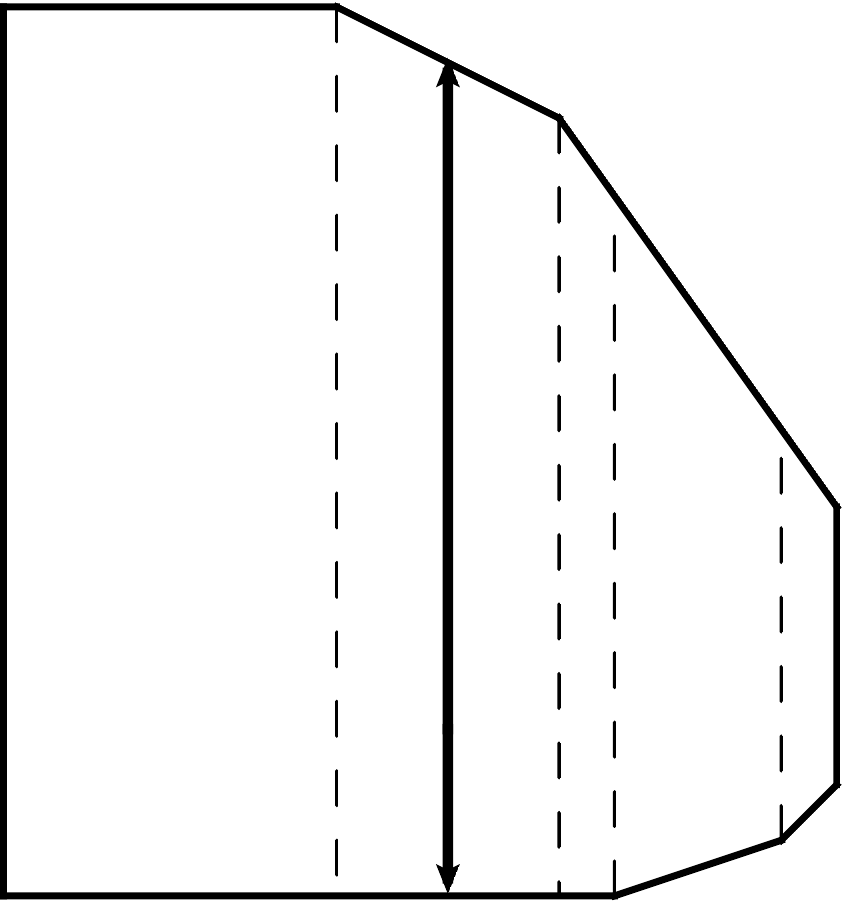}
\put (-72,50) {$\textstyle{\rho(x)}$}
\caption{Schematic cone for the $\widehat D_5$-quiver. The height of the cone gives the density $\rho(x)$ in the regions defined by the $x$ coordinates of the vertices $w_a$. }
\label{Cone}
\end{figure}

Now, let us consider the vectors $\beta_a$, $a=0,...,n+1$ as defining a set of vertices $v_a$ given by
\equn{v_a = v_0+ \textstyle{\sum_{b=0}^{a-1}}\,\beta_b\,,
}
where $v_0$ is a base point (undetermined for the moment). This set of vertices $v_a$ in turn defines a new set of edges by the equations $v_a \wedge x=1/2$. Then, the set of intersection points  of consecutive edges, given by $w_a=\beta_a/(2\, v_{a}\wedge v_{a+1})$, together with the origin defines a cone $\mathcal C$ whose area is given by
\equ{\text{Area}(\mathcal C)=\frac{1}{8} \sum_{a=0}^{n} \frac{\beta_{a+1} \wedge \beta_{a}}{(v_{a}\wedge v_{a+1})(v_{a+1}\wedge v_{a+2})}\,.
}
The denominators  $v_a \wedge v_{a+1} = v_a \wedge (v_a +\beta_a)= v_a \wedge \beta_a= \(v_0+\textstyle{\sum_{b=0}^{a-1}\beta_b}\)\wedge \beta_a$ depend on the choice of base point $v_0$. Choosing $v_0=(-n+2,-1)$, we can set $(v_{a}\wedge v_{a+1})=-1/2\, \bar \sigma_{a} $ leading to
\equ{\frac{\text{Vol}(Y)}{\text{Vol}(S^7)}=\text{Area}(\mathcal C)\,.
\label{VolP}
}
We also note that by rescaling the cone $\mathcal C$ by a factor $2 \mu$, we can actually  interpret $\rho(x)$ as the height of the cone. In Fig.~\ref{Cone} we show the rescaled cone corresponding to the $\widehat D_5$ quiver. The $x$ coordinates of the vertices of this cone correspond to the location of the kinks in $\rho(x)$ in Fig.~\ref{Ys and rho D5}. Thus, $1/2=\int dx\, \rho(x) = 4 \mu^2 \,  \text{Area}(\mathcal C)$, from where (\ref{VolP}) follows immediately.

This construction is analogous to the polygon for the $\widehat A$-quiver \cite{Gulotta:2011si}. The vectors $\beta_a$ in that case correspond to the $(1,q_a)$ charges of five-branes involved in the brane description of the theory. The addition of the two extra vectors $\beta_0$ and $\beta_{n+1}$ in the present case seems to suggest that one should also include $(0,1)$ and $(1,0)$ branes in the description of these theories.

We would like to comment that solving the matrix model under a different ordering of the $p$'s amounts to permuting them correspondingly in the expression (\ref{Dn sigmas}). Moreover, regardless of the sign and ordering of $p$'s, the denominators appearing in the expression for $\text{Vol}(Y)$ are always given by the $\bar \sigma$'s in (\ref{all sigmas}).  In the next section we will propose a general expression, which is valid for any  value of the CS levels and is explicitly invariant under Seiberg duality.

\section{\texorpdfstring{General Formula for $\widehat D_n$ Quivers}{General Formula for Dn Quivers}}\label{Free Energy for Dn}
It was shown in \cite{Willett:2011gp,Benini:2011mf} that the free energy is invariant under a generalized Seiberg duality  \cite{Aharony:1997gp,Giveon:2008zn}. For ADE quivers, Seiberg duality can be reinterpreted as the action of the Weyl group, which acts by permuting and changing the sign of an even number of $p$'s in the case of $\widehat D$-quivers. Thus, we would like to have an expression for $\text{Vol}(Y)$ that does not assume any particular relation among CS levels and is explicitly invariant under Seiberg duality. It was proposed in \cite{Gulotta:2011vp} that this can be written as a rational function whose numerator is given by $\sum_{\mathcal{R}_+} \det (\alpha_1...\alpha_n)^2 \prod _{a=1}^n |\alpha_a \cdot p |$, where $\mathcal{R}_+$ denotes all $n$-subsets of positive roots. Note that under Weyl transformations the $\bar \sigma_a$'s  defined in (\ref{all sigmas}) are simply shuffled among each other. Based on this, we propose that the general expression for the volume corresponding to $\widehat D_n$ quivers is given by
\equ{\frac{\text{Vol}(Y)}{\text{Vol}(S^7)} =\frac{\sum_{\mathcal{R}_+} \det (\alpha_1...\alpha_n)^2 \prod _{a=1}^n |\alpha_a \cdot p |}{ 2\prod_{a=0}^{n+1}\bar\sigma_a}\,.
\label{Vol D_n}
}
As we will prove below, (\ref{Vol D_n}) reduces to (\ref{Dn sigmas}) when the CS levels are ordered. 

\begin{figure}[h!]
\centering
\includegraphics[width=0.8\textwidth]{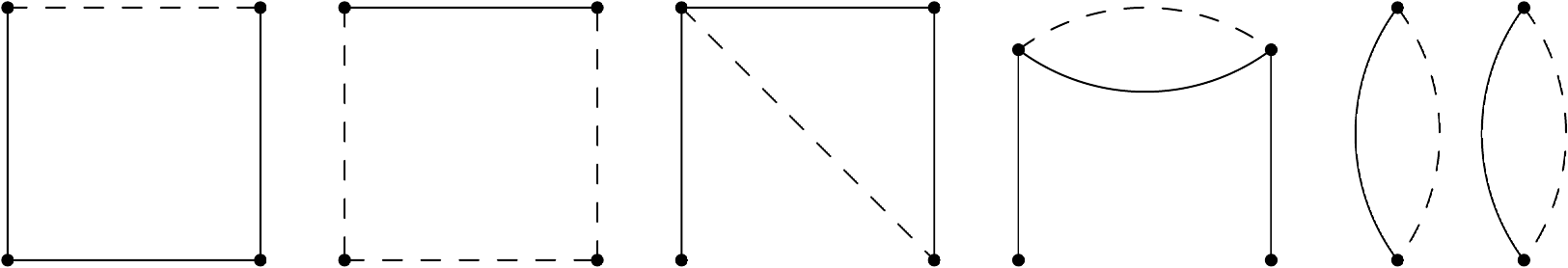}
\put (-348,30) {--} \put (-266,30) {--} \put (-176,40) {--}  \put (-104,50) {--}  \put (-44,30) {--}  \put (-14,30) {--}
\caption{Some signed graphs contributing to the numerator for $\widehat D_4$. The first diagram, for example, contributes a term $4|(p_1+p_2)(p_2-p_3)(p_3-p_4)(p_4-p_1)|$.
}
\label{Signed graphs D4}
\end{figure}
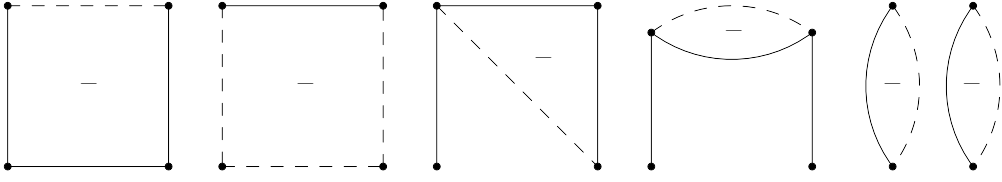

We recall that in the corresponding formula for $\widehat A$-quivers, the numerator could be interpreted as a sum over tree graphs \cite{Herzog:2010hf}.
 In a similar way, we will show now that the numerator of (\ref{Vol D_n}) can be interpreted as a sum over certain graphs known as \textit{signed graphs} \cite{SignedGraphs1} (see also \cite{SignedGraphs2, SignedGraphs3, SignedGraphs4} and references therein). A graph $\Gamma=(V,E)$ consists of a set of vertices $V$ and a set $E$ of unordered pairs from $V$ (the edges). A signed graph $(\Gamma, \sigma)$ is a graph $\Gamma$ with a signing $\sigma: \, E \rightarrow \{+1,-1\}$ associated to each edge. With these definitions, we can associate a signed graph to each term in the numerator of (\ref{Vol D_n}). Recall that the roots $\alpha_a$ for $D_n$ are of the form $\(e_i \pm e_j\)$, where $e_i$ are the canonical unit vectors of dimension $n$ and $i \neq j$. To a root  of the type $\(e_i-e_j\)$ we associate a positive edge ($\sigma=1$) connecting the nodes $i$ and $j$ in the graph, and to a root of the type $\(e_i+e_j\)$ we associate a negative edge ($\sigma=-1$). Then, we think of the matrix $I=(\alpha_1...\, \alpha_n)$ as an incidence matrix for a diagram with $n$ vertices and $n$ edges\footnote{Note that due to the absence of roots of the form $2 e_i$, one should not consider edges starting and ending on the same node.}. Due to Euler's theorem, such graphs must contain loops. If the graph contains more than one loop then it must be disconnected.  Loops are naturally associated a sign as well, given by the product of the signs of all the edges forming the loop. As we shall explain below, the determinant in (\ref{Vol D_n}) selects diagrams containing only negative loops. Some examples of diagrams contributing to the numerator for $\widehat D_4$  are shown in Fig. \ref{Signed graphs D4}, where dashed lines represent negative edges and solid lines positive ones. 

To understand why the determinant vanishes for diagrams with positive loops, it is useful to introduce the operation acting on graphs called `switching'. Switching is defined with respect to a vertex $v\in V$, and it acts by reversing the signs of all the edges connected to that vertex. This operation preserves the value of $(\det I)^2$ since it corresponds to multiplying some rows and columns of the incidence matrix $I$ by $-1$. It is easy to see that by various switching operations one can turn any loop with an even number of negative edges into a loop made entirely of positive edges. Then, $\text{det}\,I$ will vanish simply because the columns in $I$ associated to these edges add up to zero. On the other hand, if there are an odd number of negative edges in the loop, the above argument does not apply. In fact,  one can easily check that $(\det I)^2 = 4$ for each negative loop. Thus, we can also write  (\ref{Vol D_n}) as
\equ{\frac{\text{Vol}(Y)}{\text{Vol}(S^7)} =\frac{\sum_{(V,E,\sigma) ∈ \mathcal{T^-}} 4^{L_-} \prod _{(a,b) \in E} |p_a-\sigma p_b|}{2\prod_{a=0}^{n+1}\bar\sigma_a}\,,
\label{Vol D_n tree}
}
where  $\mathcal{T^-}$ denotes the set of signed diagrams with $n$ vertices and $n$ edges (connected or disconnected) and no positive loops, $L_-$ is the number of negative loops in the diagram, and $\sigma$ the sign of the corresponding edge. Using a generalized matrix-tree formula, we now show that  (\ref{Vol D_n tree}) in fact reduces to (\ref{Dn sigmas})  for $p_a>p_{a+1}$.

\subsection{Generalized Matrix-tree Formula}\label{proof}
 We define the $n \times n$ adjacency matrix $A$ for a signed graph by:  
\equn{A_{aa}= \sum_{b=1}^{n-1} (\gamma_{a,b}+\gamma_{a,-b})\, , \qquad A_{ab}=-\gamma_{a,b}+\gamma_{a,-b}\,.
}
The generalized matrix-tree formula  \cite{SignedGraphs3,SignedGraphs4}  states that 
\equ{\textstyle{\det A=\sum_{(V,E,\sigma) ∈ \mathcal{T^-}} 4^{L_-} \prod _{(a,b) \in E} |p_a-\sigma p_b|}\,.
\label{SGnumdetA}
}
By row and column operations we can bring $A$ into the tri-diagonal form: 
\equn{
A=\pmat{
\bar \sigma_1+\bar \sigma_2 + 2γ_{12} & -\bar \sigma_2 & 0 & \cdots & \cdots & 0\\
-\bar \sigma_2 & \bar \sigma_2+\bar \sigma_3 + 2γ_{23} & -\bar \sigma_3 & \cdots & \cdots & \vdots\\
0 & -\bar \sigma_3 & \ddots & \ddots & \vdots & \vdots\\
\vdots & \vdots & \ddots & \ddots & -\bar \sigma_{n-1} & 0\\
\vdots & \cdots & \cdots & -\bar \sigma_{n-1} & \bar \sigma_{n-1}+\bar \sigma_n+ 2γ_{n-1,n} & -\bar \sigma_n\\
0 & \cdots & \cdots & 0 & -\bar \sigma_n & \frac{1}{2}\(\bar \sigma_n + \bar \sigma_{n+1}\)}
}
Using the fact that the determinant of tri-diagonal matrices satisfies a recursion relation, we have
\begin{gather}
\det A=\frac{1}{2}\(\bar \sigma_n + \bar \sigma_{n+1}\)\det A_{n-1} - \bar \sigma_n^2\det A_{n-2}\,,\label{detA}\\
\det A_{a}=\(\bar \sigma_a + \bar \sigma_{a+1} + 2γ_{a,a+1}\)\det A_{a-1} - \bar \sigma_a^2\det A_{a-2}\,,\label{recdet}
\end{gather}
where $A_a$ denotes the $a\times a$ sub-matrix of $A$ for $ a=1,...,n-1$. Then, using the identities: $\bar \sigma_{a+1}-\bar \sigma_a = -2(n-2-a) \gamma_{a,a+1}$ and 
\equn{
\sum_{d=0}^{a-1}\frac{(n-2-d)γ_{d,d+1}}{\bar\sigma_d\,\bar\sigma_{d+1}}=\frac{1}{2\,\bar\sigma_a}\,,
}
we can show that the recursion relation (\ref{recdet}) is solved by
\equ{\det A_{a-1}=\prod_{b=0}^{a}\bar\sigma_b\,\sum_{d=0}^{a-1}\frac{(a-d)γ_{d,d+1}}{\bar\sigma_d\,\bar\sigma_{d+1}}\,.
\label{rec sol}
}
Using (\ref{rec sol}) in  (\ref{detA}), we have
\eqsn{\det A=&\, \frac{1}{2}\prod_{b=0}^{n+1}\bar\sigma_b\,\left[\(1+\frac{\bar\sigma_n}{\sigma_{n+1}}\)\sum_{d=0}^{n-1}\frac{(n-d)γ_{d,d+1}}{\bar\sigma_d\,\bar\sigma_{d+1}}-\frac{2\bar\sigma_n}{\bar\sigma_{n+1}}\sum_{d=0}^{n-2}\frac{(n-1-d)γ_{d,d+1}}{\bar\sigma_d\,\bar\sigma_{d+1}}\right]\\
=&\, \frac{1}{2}\prod_{b=0}^{n+1}\bar\sigma_b\,\sum_{d=0}^{n-1}\left[2\frac{γ_{d,d+1}}{\bar\sigma_d\,\bar\sigma_{d+1}}+\(\frac{\bar\sigma_{n+1}-\bar\sigma_n}{\bar\sigma_{n+1}}\)\frac{(n-2-d)γ_{d,d+1}}{\bar\sigma_d\,\bar\sigma_{d+1}}\right]\\
=&\prod_{b=0}^{n+1}\bar\sigma_b\,\left[\sum_{d=0}^{n-1}\frac{γ_{d,d+1}}{\bar\sigma_d\,\bar\sigma_{d+1}}+\frac{1}{2}\frac{4γ_{n,n+1}}{\bar\sigma_{n+1}}\frac{1}{2\bar\sigma_n}\right]\\
=&\prod_{b=0}^{n+1}\bar\sigma_b\,\sum_{d=0}^{n}\frac{γ_{d,d+1}}{\bar\sigma_d\,\bar\sigma_{d+1}}\,.
}
Finally, substituting (\ref{SGnumdetA}) in (\ref{Vol D_n tree}) leads to
\equn{\frac{\text{Vol}(Y)}{\text{Vol}(S^7)}=\frac{\det A}{2\prod_{b=0}^{n+1}\bar\sigma_b}=\frac{1}{2}\sum_{d=0}^{n}\frac{γ_{d,d+1}}{\bar\sigma_d\,\bar\sigma_{d+1}} \,,
}
recovering the expression (\ref{Dn sigmas}).

\section{\texorpdfstring{Flavored $\widehat D_n$ Quivers and the F-theorem}{Flavored Dn Quivers and the F-theorem}}\label{Flavored $D_n$ quivers and the F-theorem}
The F-Theorem \cite{Jafferis:2011zi} states that the free energy (\ref{F log Z})  decreases along RG flows and is stationary at the RG fixed points of any three-dimensional field theory (supersymmetric or not). Thus, $F$ gives a good measure of the number of degrees of freedom, in analogy with the c-function in two dimensions and the anomaly coefficient, $a$ in four dimensions. This theorem was first tested in a variety of field theories \cite{Klebanov:2011gs, Amariti:2011da, Klebanov:2011td} and recently it has been proven in \cite{Casini:2011kv, Casini:2012ei} for any three-dimensional field theory by relating $F$ to the entanglement entropy of a disk-like region. Here we check that it holds for the the class of theories we have discussed. We trigger the RG flow by  adding massive non-chiral fundamental flavors in the UV. By integrating out non-chiral flavor fields, there is no effective shift in the CS levels. Thus, we are interested in comparing $F(k_i;n_F)$ to $F(k_i;0)$. The addition of $n_F \neq 0$ in (\ref{F}) introduces no additional complications and the matrix model is solved as explained in section \ref{Solving the Matrix Models}. We solved the flavored $\widehat D_n$ matrix model for $n=4,...,9$ leading us to
\equ{\frac{\text{Vol}(Y)}{\text{Vol}(S^7)}=\frac{1}{2} \(\frac{\gamma_{01}}{\bar \sigma_0(\bar \sigma_1+n_F)} + \sum_{a=1}^{n} \frac{\gamma_{a,a+1}}{(\bar \sigma_a+n_F) (\bar \sigma_{a+1}+n_F)}\).
\label{Partial fractions Dn nf}
}
By comparing (\ref{Partial fractions Dn nf}) with (\ref{Dn sigmas}), it is clear that $F(k_i; n_F) \geq F(k_i; 0)$ verifying that
\equn{F_{UV} \geq F_{IR}\,,}
in accordance with the F-theorem. 

In terms of the polygon construction discussed in Section~\ref{General solution and polygon area}, adding flavor corresponds to adding the vector $\beta_F=(0, n_F/2)$ between $\beta_0$ and $\beta_1$. Then, (\ref{Partial fractions Dn nf}) has the same form as (\ref{Dn sigmas}) with $b=F,1,...,n$ in the definition (\ref{all sigmas}).

\section{\texorpdfstring{Unfolding $\widehat D_n$ to $\widehat A_{2n-5}$}{Unfolding Dn to A(2n-5)}}\label{Unfolding section}
Here we provide a check of the formula (\ref{Dn sigmas}), based on the folding$/$unfolding trick discussed in \cite{Gulotta:2012yd}, which relates the free energy of various quiver gauge theories when some CS levels are identified. It can be used to change the gauge groups from unitary to orthosymplectic without changing the quiver or it can be used to change the quiver without changing the type of gauge group. Here we will deal with the latter use, as it relates the free energy of $\widehat D$-quivers to that of $\widehat A$-quivers.

\begin{figure}[h]
\centering
\begin{tabular}{ccc}
\includegraphics[width=2.5in]{Dn.pdf} & $\qquad\qquad$ &\includegraphics[width=2.5in]{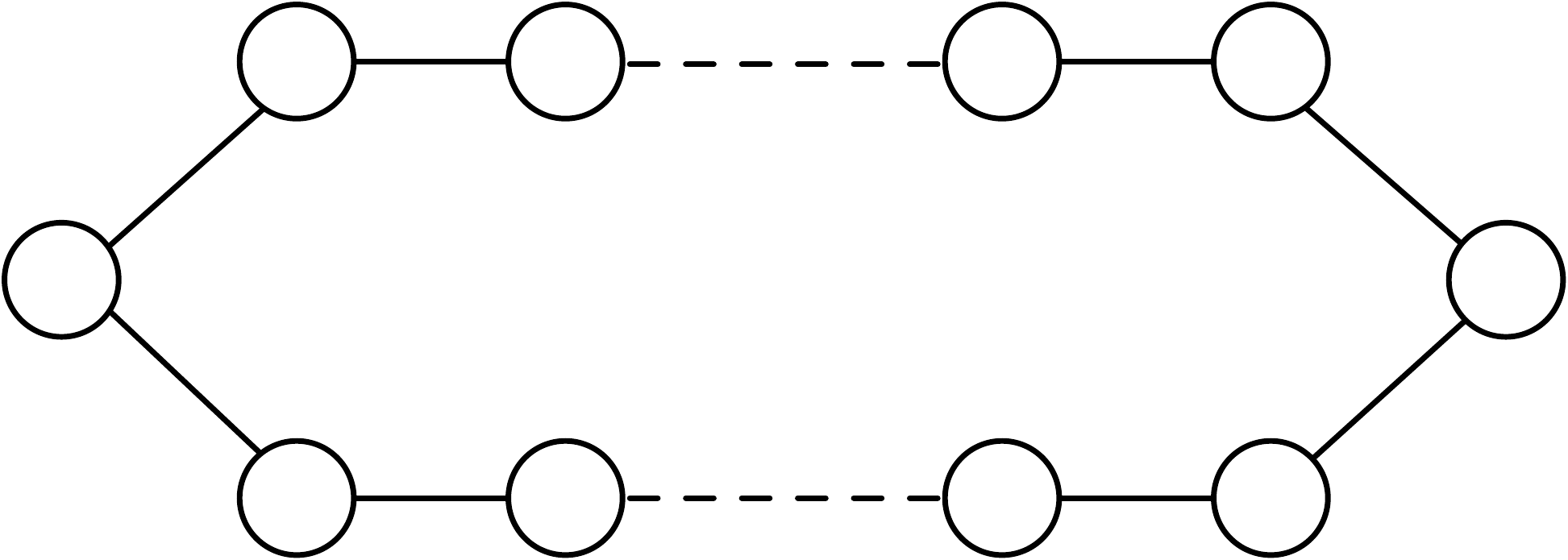}
\put(-240,28){$\xrightarrow{\mathit{Unfolding}}$}
\put (-260,-10) {$k'$} \put (-260,67) {$k'$}\put (-430,67) {$k$} \put (-430,-10) {$k$} \put (-400,42) {$k_5$} \put (-370,42) {$k_6$} \put (-345,42) {$\cdots$} \put (-303,42) {$k_{n+1}$}
\put (-12,41) {$2k'$} \put (-181,16) {$2k$} \put (-150,-10) {$k_5$} \put (-120,-10) {$k_6$} \put (-95,-10) {$\cdots$} \put (-50,-9) {$k_{n+1}$} \put (-150,67) {$k_5$} \put (-120,67) {$k_6$} \put (-95,67) {$\cdots$} \put (-50,67) {$k_{n+1}$}
\end{tabular}
\caption{Unfolding $\widehat D_n$ to $\widehat A_{2n-5}$. Each node in the $\widehat A$ quiver corresponds to a $U(2N)$ gauge group.}
\label{unfolding}
\end{figure}

When the external CS levels of a $\widehat D_n$ quiver are identified, it can be unfolded to an $\widehat A_{2n-5}$ quiver, as shown in Fig.~\ref{unfolding}. Each internal node in the $\widehat D$ quiver is duplicated to give two nodes with the same CS level, while the four external nodes combine to give two nodes with doubled CS levels. Each node in the $\widehat A$ quiver corresponds to a $U(2N)$ gauge group and the condition $\sum_a n_a k_a=0$ is automatically satisfied in the unfolded quiver. Using this, it can be shown that in the large $N$ limit,  $Z_D=\sqrt{Z_A}$ and therefore the free energies are related by $F_D=\frac{1}{2}F_A$. Here we verify explicitly this proportionality by comparing the formula (\ref{Dn sigmas}) to the corresponding formula for $\widehat A_{2n-5}$.

Let us first look at the formula for the $\widehat D_n$ quiver when external CS levels are identified, $i.e.$, $k_1=k_2=k$ and $k_3 =k_4 = k'$. Due to the relations in (\ref{k2p}), this is ensured by setting $p_1=p_n=0$.  Thus, we need the solution to the matrix model with the ordering $p_2 \geq  ...\geq p_{n-1} \geq p_n \geq p_1 \geq 0$. As mentioned at the end of Section~\ref{Solving the Matrix Models}, this is given by permuting the $p$'s in (\ref{Dn sigmas}) accordingly. Then,  setting $p_{1}=p_n=0$ gives
\equ{\frac{\text{Vol}(Y_D)}{\text{Vol}(S^7)}=\frac{1}{2} \(\frac{p_2}{\bar \sigma_2^2} + \sum_{a=2}^{n-2} \frac{\gamma_{a,a+1}}{\bar \sigma_a\, \bar \sigma_{a+1}}  +\frac{p_{n-1}}{ (\bar \sigma_{n-1})^2 }\).
\label{Volume D folded}
}

Now we wish to compare this expression with the corresponding one for $\widehat A_{2n-5}$ \cite{Gulotta:2011si}, namely
\equ{\frac{\text{Vol}(Y_A)}{\text{Vol}(S^7)}=\frac{1}{2} \sum_{a=1}^{2n-4} \frac{\gamma_{a,a+1}}{\sigma_a\, \sigma_{a+1}}\,,
\label{Partial fractions An}
}
where $\sigma_a = \sum_{a=1}^{2n-4}|q_a-q_b|$, $\gamma_{a,b}= |q_a-q_b|$ and $\sum_{a=1}^{2n-4}q_a=0$. The identification of opposite CS levels in the $\widehat A_{2n-4}$ quiver leads to $q_a=-q_{2n-3-a}$ (see Appendix~\ref{A and D Appendix} for details). Then, we assume that $q_{1} \geq ... \geq q_{n-2} \geq 0 \geq q_{n-1} \geq ... \geq q_{2n-4}$ and
\[\sigma_a=\sum_{b=1}^{n-2}|q_a-q_b|+\sum_{b=n-1}^{2n-4}|q_a-q_b|=\sum_{b=1}^{n-2}(|q_a-q_b|+|q_a+q_b|)\,; \quad a=1,...,n-2\,.\]
Noting that $\sigma_a =\bar\sigma_a$ and $q_a=p_{a+1}$ for $a=1,...,n-2$, we have
\eqs{\frac{\text{Vol}(Y_A)}{\text{Vol}(S^7)}=&\,\frac{1}{2} \left( \sum_{a=1}^{n-3} \frac{\gamma_{a,a+1}}{\sigma_a\, \sigma_{a+1}} + \frac{\gamma_{n-2,n-1}}{\sigma_{n-2}\,\sigma_{n-1}} + \sum_{a=n-1}^{2n-5} \frac{\gamma_{a,a+1}}{\sigma_a\,\sigma_{a+1}} + \frac{\gamma_{2n-4,2n-3}}{\sigma_{2n-4}\,\sigma_{2n-3}} \right)\nonumber\\
=&\sum_{a=1}^{n-3} \frac{\gamma_{a,a+1}}{\bar \sigma_a\,\bar \sigma_{a+1}} + \frac{q_{n-2}}{\(\bar \sigma_{n-2}\)^2} + \frac{q_1}{\bar \sigma_{1}^2}\nonumber\\
=&\,\frac{p_2}{\bar \sigma_2^2} + \sum_{a=2}^{n-2} \frac{\gamma_{a,a+1}}{\bar \sigma_a\, \bar \sigma_{a+1}}  +\frac{p_{n-1}}{ (\bar \sigma_{n-1})^2}\,.\label{Volume A folded}
}
Thus, comparing (\ref{Volume A folded}) to (\ref{Volume D folded}) we have
\equ{ \text{Vol}(Y_D)  = \frac{1}{2}\,\text{Vol}(Y_A)\,.
}

\begin{figure}[h]
\centering
\begin{tabular}{VVV}
\includegraphics[width=1in,height=2in]{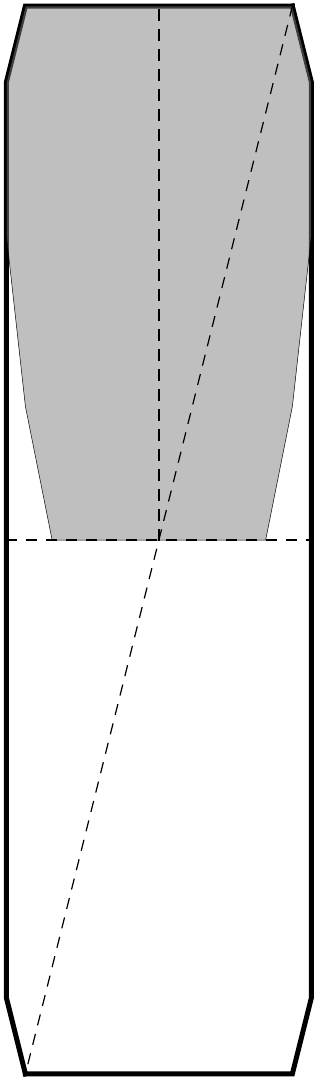} & $\xrightarrow{\mathit{Unfolding}}$ &\includegraphics[width=1in,height=2in]{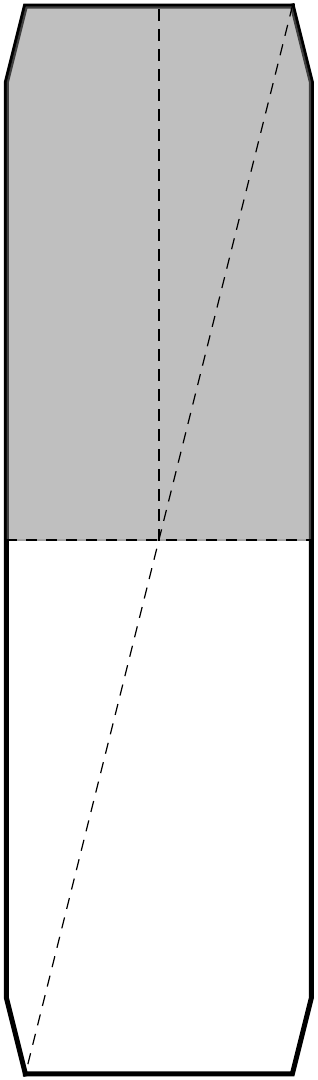}
\end{tabular}
\caption{Polygons associated to the $\widehat D_4$ quiver (shaded region) and $\widehat A_3$ quiver (outer polygon). Upon unfolding, $\text{Area}(\mathcal P_D)=1/2\, \text{Area}(\mathcal P_A)$.}
\label{FoldingGeo}
\end{figure}

This relation can also be seen clearly in terms of the areas of the corresponding polygons, as shown in Fig.~\ref{FoldingGeo} (the cone as defined in Section \ref{General solution and polygon area} has been doubled along the dotted line for visual clarity). The outer polygon corresponds to the $\widehat A$-quiver with opposite CS levels identified and the shaded region on the left represents the polygon corresponding to a general $\widehat D$-quiver; when $p_1=p_n=0$, this shaded region expands to fill half of the outer polygon on the right.

Recalling that the nodes of the unfolded $\widehat A$-quiver correspond to $U(2N)$ gauge groups, we verify that
\equn{\frac{F_{D}}{F_A}=\frac{N^{3/2}}{(2N)^{3/2}}\sqrt{\frac{\text{Vol}(Y_A)}{\text{Vol}(Y_D)}}=\frac{1}{2}\,.}

\vspace{10pt}
\section{Discussion}
In this paper we have studied  three-dimensional $\widehat D_n$ quiver Chern-Simons matter theories  by using the localization method of Kapustin, Willet and Yaakov in the large $N$ limit. These field theories are believed to be dual to M-theory on $AdS_4 \times Y$, where $Y$ is a tri-Sasaki Einstein manifold. We have explicitly solved  the corresponding matrix models for various values of $n$, leading us to conjecture a general expression for the free energy and therefore for the volume of the corresponding space $Y$ given in (\ref{Vol D_n}). We have shown that the numerator of this expression can be interpreted as a sum over a class of graphs  with edges that carry a sign, known as signed graphs.  Using a generalized matrix-tree formula, we prove that for a particular ordering of CS levels, it can also be interpreted as  the area of a certain polygon, given by (\ref{Dn sigmas}). When external CS levels  in the $\widehat D_n$ quiver are identified, the area of this polygon becomes half the area of the polygon corresponding to the $\widehat A_{2n-5}$ quiver, in accordance with the unfolding procedure.  We have also studied the addition of massive flavor fields, showing that when they are integrated out, the area of the corresponding polygon always increases (thereby decreasing $F$), in accordance with the F-theorem. 

The relevant tri-Sasaki Einstein space for a $\widehat D_n$ quiver is the base of the hyperk\"{a}hler cone defined by the quotient $\mathbb H^{4n-8}/// U(1)^{n-1} \times SU(2)^{n-3}$. To the best of our knowledge, the volumes of these spaces have not been computed. Thus, (\ref{Vol D_n}) can be considered as an AdS/CFT prediction for these volumes.  A possible approach to proving the conjectured expression for the free energy would be to find the general solution to the matrix model, perhaps in terms of the polygon construction presented above, as it has been done for the $\widehat A$-quiver in \cite{Gulotta:2011si}.  Some questions which have not been addressed here are whether there is a  group theory interpretation of the volume formula and whether its denominator can be written in a form that is universal for any ADE quiver.

\vspace{-30pt}
\section*{\cen{Acknowledgements}}
We would like to thank Daniel Gulotta for various discussions and clarifications. This  work is supported in part by NSF grants PHY-0969739, PHY-0844827 and PHY-0756966. C.H. also thanks the Sloan Foundation for partial support.

\appendix
\section{\texorpdfstring{Roots of $\widehat A_{m-1}$ and $\widehat D_n$}{Roots of A(m-1) and Dn}}\label{A and D Appendix}
Here we  give some useful information about the roots for $\widehat A$ and $\widehat D$ Lie algebras. For $\widehat A_{m-1}$ we choose the following root basis
\equn{\tilde \alpha_a=e_a- e_{a+1}\,, \quad a=1,...,m-1\,; \qquad \tilde \theta=- e_1+ e_m\,,
}
where $e_a$  are canonical unit vectors of dimension $m$. For $\widehat D_n$ we choose
\equn{ \alpha_i =  e_i- e_{i+1}\,, \quad i=1,...,n-1\,; \qquad  \alpha_n= e_{n-1}+e_n\,, \qquad  \theta = -(e_1+e_2)\,,
}
where $e_i$ are the unit vectors of dimension $n$.
\begin{figure}[h]
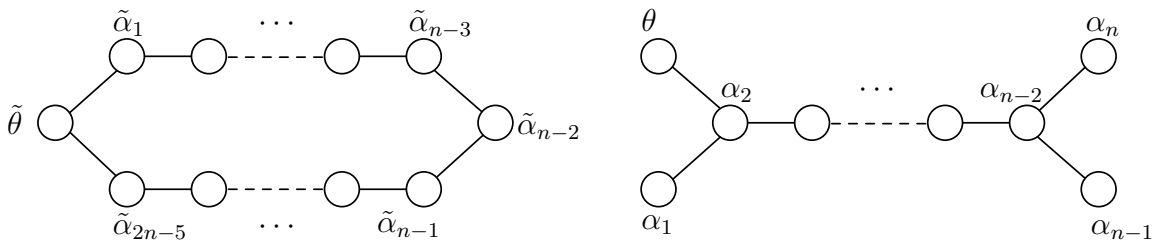

\centering
\begin{tabular}{ccc}
\includegraphics[width=2.5in]{An.pdf} & \qquad\qquad &\includegraphics[width=2.5in]{Dn.pdf}
\put (-227,28) {$\tilde{\alpha}_{n-2}$} \put (-420,28) {$\tilde \theta$} \put (-380,-10) {$\tilde{α}_{2n-5}$} \put (-325,-10) {$\cdots$} \put (-280,-9) {$\tilde{α}_{n-1}$} \put (-380,67) {$\tilde{α}_{1}$} \put (-325,67) {$\cdots$} \put (-268,67) {$\tilde{α}_{n-3}$}
\put (-10,-10) {$\alpha_{n-1}$} \put (-12,67) {$α_n$}\put (-180,67) {$\theta$} \put (-180,-8) {$α_1$} \put (-150,42) {$α_{2}$} \put (-98,42) {$\cdots$} \put (-52,42) {$α_{n-2}$}
 \end{tabular}
\caption{Dynkin diagrams for $\widehat A_{2n-5}$ and $\widehat D_{n}$.}
\label{A and D quivers}
\end{figure}

In Fig.~\ref{A and D quivers} we show the affine Dynkin diagrams for the $\widehat A$ and $\widehat D$ Lie algebras along with the roots associated with every node. At each node, the CS level is given by $\tilde \alpha \cdot q$ and $\alpha \cdot p$ for $\widehat A$ and $\widehat D$, respectively. The identification of opposite CS levels in the $\widehat A_{2n-5}$ quiver imposes $\tilde \alpha_a\cdot q=\tilde \alpha_{2n-4-a} \cdot q$ and hence $q_{a}=-q_{2n-3-a}$ for $a=1,...,n-2$. With these conventions, unfolding the $\widehat D$-quiver to the $\widehat A$-quiver relates $q_a=p_{a+1}$.

\section{\texorpdfstring{$\widehat D_5$}{D₅}}\label{D5App}
Here we give the detailed solution of the matrix model for the $\widehat D₅$ quiver gauge theory. As discussed in Section \ref{Solving the Matrix Models}, there are 7 regions defining a generic solution of this model. To keep the notation simple, the second index for the four $y$'s corresponding to the external nodes  is suppressed. 
\subsubsection*{Region 1: $0\leq x\leq \frac{\mu }{3(k_2+k_3+k_4+2 k_5+2 k_6)}$}
\begin{gather*}
\rho=\frac{\mu}{3}\,;\\
y_1-y_{6,2}=\frac{\(2k_1-k_3-k_4-2k_6\)x}{4\rho}\,,\quad y_2-y_{6,2}=\frac{\(2k_2-k_3-k_4-2k_6\)x}{4\rho}\,,\quad y_3-y_{6,2}=\frac{k_3 x}{2\rho}\,,\\
y_4-y_{6,2}=\frac{k_4 x}{2\rho}\,,\quad y_{5,1}-y_{6,2}=y_{5,2}-y_{6,2}=-\frac{(k_3+k_4+2k_6)x}{4\rho}\,,\quad y_{6,1}-y_{6,2}=0\,.
\end{gather*}

\subsubsection*{Region 2: $\frac{\mu }{3(k_2+k_3+k_4+2 k_5+2 k_6)}\leq x\leq \frac{4\mu}{6k_2+9k_3+9k_4+12k_5+18k_6}$}
\begin{gather*}
\rho=\frac{\mu}{3}\,;\\
y_1-y_{5,2}=-\frac{1}{2}\,,\quad y_2-y_{6,2}=\frac{\(2k_2-k_3-k_4-2k_6\)x}{4\rho}\,,\quad y_3-y_{6,2}=\frac{k_3 x}{2\rho}\,,\\
y_4-y_{6,2}=\frac{k_4 x}{2\rho}\,,\quad y_{5,1}-y_{6,2}=-\frac{1}{2}-\frac{\(2k_1+k_3+k_4+2k_6\)x}{4\rho}\,,\\
y_{5,2}-y_{6,2}=\frac{1}{2}+\frac{\(2k_1-k_3-k_4-2k_6\)x}{4\rho}\,,\quad y_{6,1}-y_{6,2}=0\,.
 \end{gather*}

\subsubsection*{Region 3:  $\frac{4\mu}{6k_2+9k_3+9k_4+12k_5+18k_6}\leq x\leq \frac{2\mu }{3(2k_2+k_3+k_4+2k_5+2k_6)}$}
\begin{gather*}
\rho=\frac{\mu}{3}\,;\\
y_1-y_{5,2}=-\frac{1}{2}\,,\quad y_2-y_{6,2}=-1-\frac{\(k_1-k_2\)x}{2\rho}\,,\quad y_3-y_{6,2}=-1-\frac{\(2k_1-3k_3-k_4-2k_6\)x}{4\rho}\,,\\
y_4-y_{6,2}=-1-\frac{\(2k_1-k_3-3k_4-2k_6\)x}{4\rho}\,,\quad y_{5,1}-y_{6,2}=-\frac{3}{2}-\frac{k_1 x}{\rho}\,,\\
y_{5,2}-y_{6,2}=-\frac{1}{2}\,,\quad y_{6,1}-y_{6,2}=-2-\frac{\(2k_1-k_3-k_4-2k_6\)x}{2ρ}\,.
 \end{gather*}

\subsubsection*{Region 4: $\frac{2\mu }{3(2k_2+k_3+k_4+2k_5+2k_6)}\leq x\leq \frac{2\mu }{2k_2+3(k_3+k_4+2k_5+2k_6)}$}
\begin{gather*}
\rho=\frac{\mu}{2}+\frac{x}{4}\(k_1-k_2\)\,;\\
y_1-y_{5,2}=-\frac{1}{2}\,,\quad y_2-y_{6,2}=0\,,\quad y_3-y_{6,2}=-\frac{1}{2}+\frac{\(2k_3+k_4+k_5+2k_6\)x}{2\rho}\,,\\
y_4-y_{6,2}=-\frac{1}{2}+\frac{\(k_3+2k_4+k_5+2k_6\)x}{2\rho}\,,\quad y_{5,1}-y_{6,2}=-\frac{1}{2}-\frac{\(k_1-k_2\)x}{2\rho}\,,\\
y_{5,2}-y_{6,2}=-\frac{1}{2}\,,\quad y_{6,1}-y_{6,2}=-1+\frac{\(k_3+k_4+k_5+2k_6\)x}{ρ}\,.
 \end{gather*}

\subsubsection*{Region 5: $\frac{2\mu }{2k_2+3(k_3+k_4+2k_5+2k_6)}\leq x\leq \frac{2\mu }{2k_2+3k_3+5k_4+4k_5+6k_6}$}
\begin{gather*}
\rho=\mu+x k_1\,;\\
y_1-y_{5,2}=-\frac{1}{2}\,,\quad y_2-y_{6,2}=0\,,\quad y_3-y_{6,2}=\frac{\(3k_3+k_4+2k_6\)x}{4\rho}\,,\\
y_4-y_{6,2}=\frac{\(k_3+3k_4+2k_6\)x}{4\rho}\,,\quad y_{5,1}-y_{6,2}=\frac{1}{2}\,,\\
y_{5,2}-y_{6,2}=-\frac{1}{2}\,,\quad y_{6,1}-y_{6,2}=\frac{\(k_3+k_4+2k_6\)x}{2ρ}\,.
 \end{gather*}

\subsubsection*{Region 6: $\frac{2\mu }{2k_2+3k_3+5k_4+4k_5+6k_6}\leq x\leq \frac{2\mu }{2k_2+5k_3+3k_4+4k_5+6k_6}$}
\begin{gather*}
\rho=\frac{3\mu}{2}+\frac{x}{4}\(6k_1-k_3-3k_4-2k_6\)\,;\\
y_1-y_{5,2}=-\frac{1}{2}\,,\quad y_2-y_{6,2}=0\,,\quad y_3-y_{6,2}=\frac{1}{6}+\frac{\(2k_3+k_6\)x}{3\rho}\,,\\
y_4-y_{6,2}=\frac{1}{2}\,,\quad y_{5,1}-y_{6,2}=\frac{1}{2}\,,\quad y_{5,2}-y_{6,2}=-\frac{1}{2}\,,\quad y_{6,1}-y_{6,2}=\frac{1}{3}+\frac{\(k_3+2k_6\)x}{3ρ}\,.
 \end{gather*}

\subsubsection*{Region 7:  $\frac{2\mu }{2k_2+5k_3+3k_4+4k_5+6k_6}\leq x\leq \frac{2\mu }{2k_2+3k_3+3k_4+4k_5+6k_6}$}
\begin{gather*}
\rho=2\mu+x\(2k_1-k_3-k_4-k_6\)\,;\\
y_1-y_{5,2}=-\frac{1}{2}\,,\quad y_2-y_{6,2}=0\,,\quad y_3-y_{6,2}=\frac{1}{2}\,,\quad y_4-y_{6,2}=\frac{1}{2}\,,\\
y_{5,1}-y_{6,2}=\frac{1}{2}\,,\quad y_{5,2}-y_{6,2}=-\frac{1}{2}\,,\quad y_{6,1}-y_{6,2}=\frac{1}{2}+\frac{k_6 x}{2ρ}\,.
 \end{gather*}
Finally, the last saturation occurs at the end of this region with $y_{6,1}=y_{6,2}+1$.

\section{Exceptional Quivers}
\label{Exceptional Quivers}
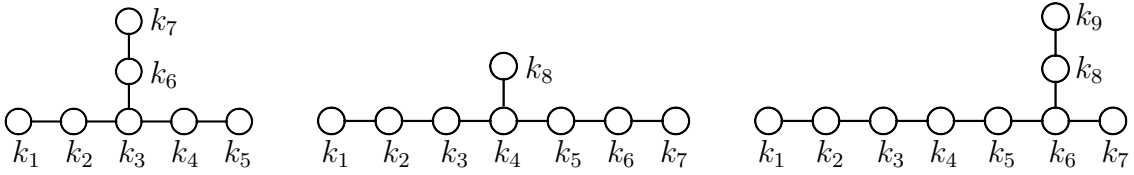
\begin{figure}[h]
\centering
\begin{tabular}{ccc}
\includegraphics[width=0.2\textwidth ]{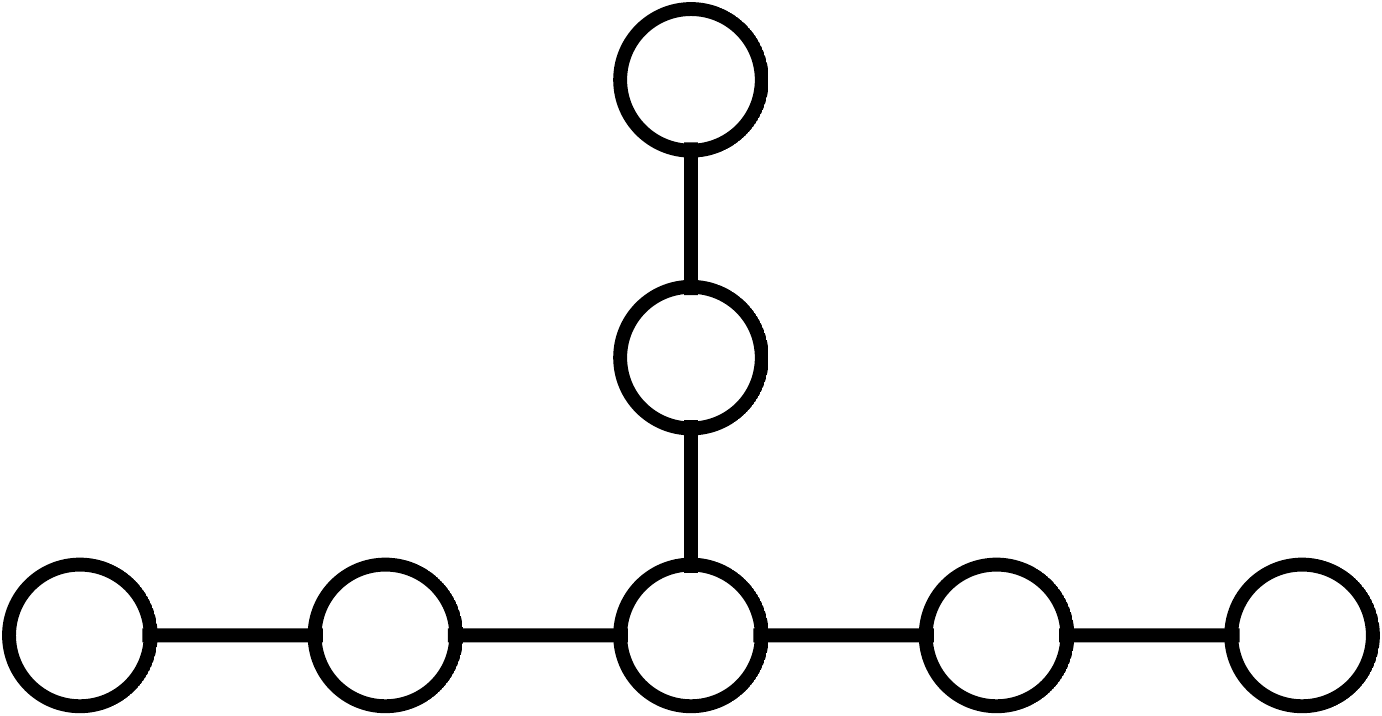} & \quad \includegraphics[width=0.3\textwidth]{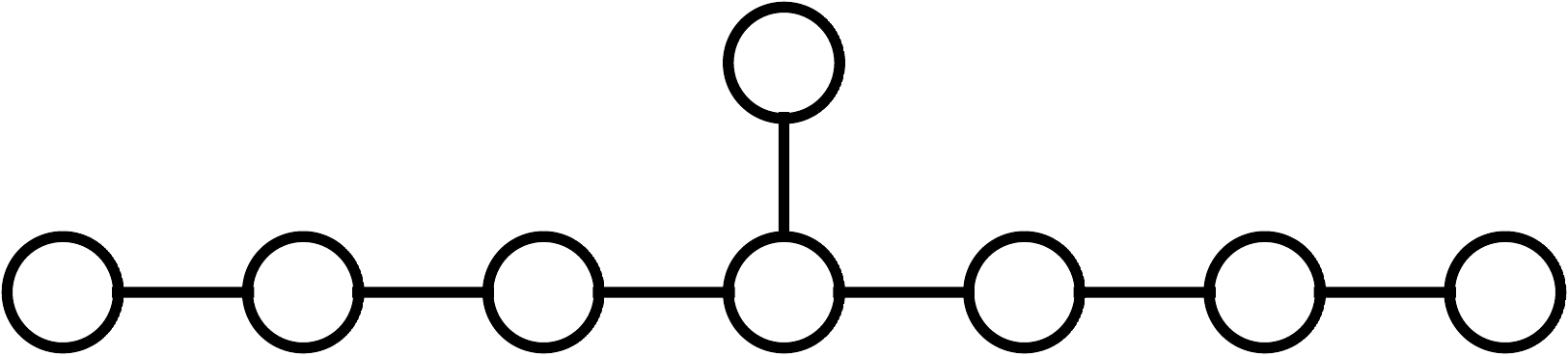}  & \quad \includegraphics[width=0.3\textwidth]{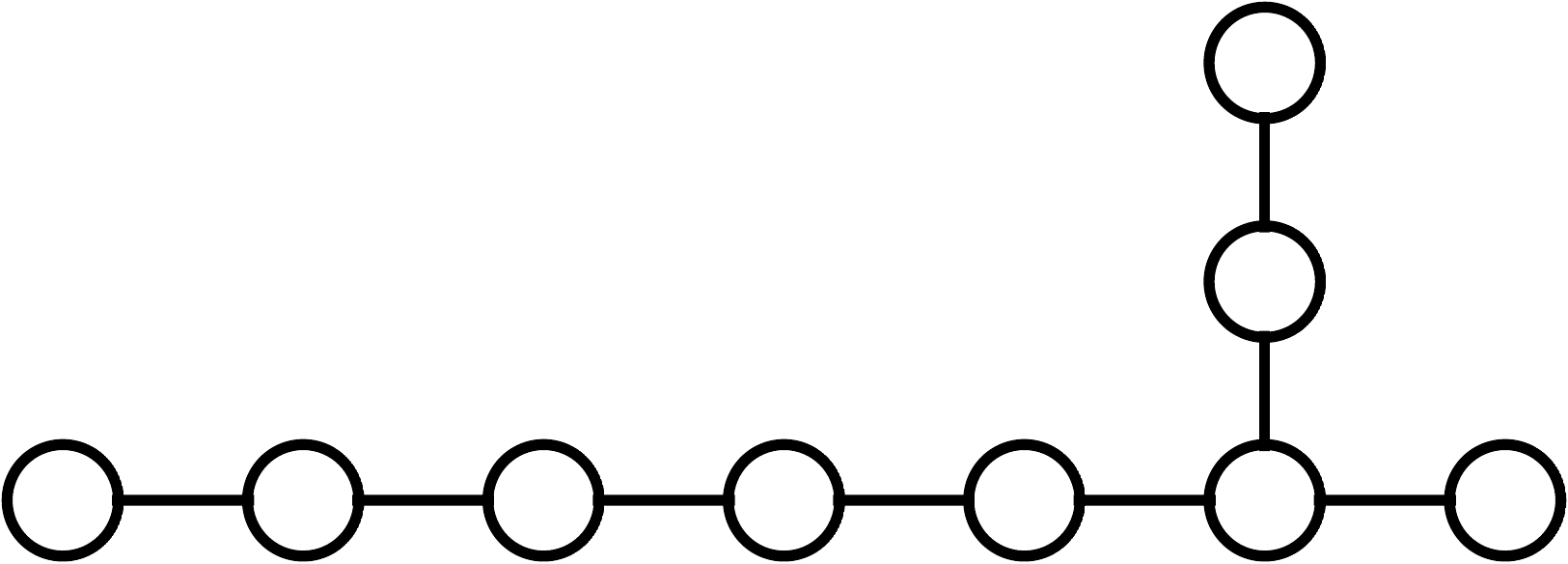} 
\put(-422,-10) {$k_1$} \put(-402,-10) {$k_2$} \put(-382,-10) {$k_3$} \put(-362,-10) {$k_4$} \put(-342,-10) {$k_5$}   \put(-370,20) {$k_6$} \put(-370,40) {$k_7$} 
\put(-305,-10) {$k_1$} \put(-282,-10) {$k_2$} \put(-260,-10) {$k_3$} \put(-240,-10) {$k_4$} \put(-217,-10) {$k_5$}  \put(-197,-10) {$k_6$} \put(-177,-10) {$k_7$}  \put(-228,22) {$k_8$}
\put(-140,-10) {$k_1$} \put(-118,-10) {$k_2$} \put(-95,-10) {$k_3$} \put(-75,-10) {$k_4$} \put(-53,-10) {$k_5$}  \put(-30,-10) {$k_6$} \put(-10,-10) {$k_7$}   \put(-20,22) {$k_8$}  \put(-20,42) {$k_9$}
\end{tabular}
\caption{Labeling of Chern-Simons levels for $\widehat E_6, \widehat E_7$ and $\widehat E_8$.}
\label{E's}
\end{figure}

We have also solved the matrix models for the exceptional quivers $\widehat E_6, \widehat E_7$ and $\widehat E_8$. They consist of eleven, seventeen and twenty-nine regions respectively. Here we give the corresponding free energies for a particular ordering of the CS levels. In Fig.~\ref{E's}, we show our conventions in labeling the nodes.
 
\subsection*{$\widehat E_6$}
The matrix model for $\widehat E_6$ gives:
\eqsn{\frac{2}{\mu^2}=&\frac{2 (4 k_2+11 k_3+8 k_4+4 k_5+6 k_6+4 k_7)}{(2 k_2+5 k_3+4 k_4+2 k_5+3 k_6+2 k_7)^2}-\frac{1}{42 (3 k_2+6 k_3+4 k_4+2 k_5+5 k_6+k_7)}\\
&-\frac{1}{77 (13 k_2+12 k_3+8 k_4+4 k_5+3 k_6+2 k_7)}-\frac{1}{3 (3 k_2+6 k_3+4 k_4+2 k_5+5 k_6+4 k_7)}\\
&-\frac{9}{6 k_2+14 k_3+13 k_4+6 k_5+9 k_6+6 k_7}-\frac{9}{11 (6 k_2+14 k_3+13 k_4+12 k_5+9 k_6+6 k_7)}\,,
}
under the assumptions that $k_6 \geq k_5 \geq k_4 \geq k_3 \geq k_2 \geq 0$ and $k_7 > 3 k_2 + 6 k_3 + 4 k_4 + 2 k_5 + 4 k_6$. 

\subsection*{$\widehat E_7$}
The matrix model for $\widehat E_7$ gives:
\allowdisplaybreaks{\eqsn{\frac{2}{\mu^2}=&\frac{8 k_2+24 k_3+42 k_4+4 (8 k_5+6 k_6+3 k_7+5 k_8)}{(2 k_2+6 k_3+10 k_4+8 k_5+6 k_6+3 k_7+5 k_8)^2}\\
&-\frac{1}{2 k_2+7 k_3+10 k_4+8 k_5+6 k_6+3 k_7+5 k_8}-\frac{1}{2 k_2+6 k_3+10 k_4+9 k_5+6 k_6+3 k_7+5 k_8}\\
&-\frac{1}{180 (2 k_2+3 k_3+4 k_4+3 k_5+2 k_6+k_7+2 k_8)}\\
&-\frac{4}{15 (4 k_2+11 k_3+2 (9 k_4+8 k_5+7 k_6+6 k_7)+9 k_8)}\\
&-\frac{27}{7 (6 k_2+17 k_3+28 k_4+24 k_5+20 k_6+9 k_7+15 k_8)}\\
&-\frac{32}{21 (8 k_2+25 k_3+42 k_4+32 k_5+22 k_6+12 k_7+27 k_8)}\,,
}}

\noindent under the assumptions that $k_7 \geq k_6 \geq k_5 \geq k_4 \geq k_3 \geq k_2 \geq 0$ along with $4 k_3 + k_4 > 2 k_5 + k_6\,, k_3 + 2 k_4 + k_5 > k_7\,, k_4 + k_5 > k_6\,\,\text{and}\,\,3 k_8 > 6 k_3 + 12 k_4 + 15 k_5 + 10 k_6 + 5 k_7$.

\subsection*{$\widehat E_8$}
The solution of the matrix model for $\widehat E_8$ gives:
\allowdisplaybreaks{\eqsn{\frac{2}{\mu^2}=&\frac{8 k_2+24 k_3+48 k_4+74 k_5+92 k_6+48 k_7+64 k_8+32 k_9}{(2 k_2+6 k_3+12 k_4+18 k_5+23 k_6+12 k_7+16 k_8+8 k_9)^2}\\
&-\frac{1}{3150 (2 k_2+3 k_3+4 k_4+5 k_5+6 k_6+3 k_7+4 k_8+2 k_9)}\\
&-\frac{1}{2 (k_2+3 k_3+6 k_4+9 k_5+12 k_6+6 k_7+8 k_8+4 k_9)}\\
&-\frac{1}{2 k_2+6 k_3+13 k_4+18 k_5+23 k_6+12 k_7+16 k_8+8 k_9}\\
&-\frac{27}{7 (6 k_2+18 k_3+35 k_4+52 k_5+69 k_6+38 k_7+48 k_8+24 k_9)}\\
&-\frac{108}{35 (12 k_2+36 k_3+70 k_4+104 k_5+138 k_6+69 k_7+103 k_8+48 k_9)}\\
&-\frac{36}{55 (12 k_2+36 k_3+70 k_4+104 k_5+138 k_6+69 k_7+103 k_8+68 k_9)}\\
&-\frac{9}{154 (6 k_2+17 (3 k_3+4 k_4+5 k_5+6 k_6+3 k_7+4 k_8+2 k_9))}\,,
}}

\noindent assuming that $k_6 \geq k_5 \geq k_4 \geq k_3 \geq k_2 \geq 0\,, k_7 > 3 k_4 + 6 k_5 + 4 k_6\,, 2 k_4 + 4 k_5 + 6 k_6 + 9 k_7 > k_8\,, 2 k_3 + 4 k_4 + 6 k_5 + 8 k_6 + 4 k_7 + 6 k_8 > k_9\,\,\text{and}\,\,2 k_9 > 6 k_3 + 12 k_4 + 18 k_5 + 24 k_6 + 16 k_7 + 11 k_8$.

\references{
\bibitem{Witten:1988ze}
E. Witten,
``Topological Quantum Field Theory'',
\cmp{117}{1988}{353}.

\bibitem{Pestun:2007rz}
V. Pestun,
``Localization of Gauge Theory on a Four-sphere and Supersymmetric Wilson Loops'',
\cmp{313}{2012}{71} [\arXivid{0712.2824} {\color{cyan}\small [hep-th]}].

\bibitem{Kapustin:2009kz}
A. Kapustin, B. Willett and I. Yaakov,
``Exact Results for Wilson Loops in Superconformal Chern-Simons Theories with Matter'',
\jhep{1003}{2010}{089} [\arXivid{0909.4559} {\color{cyan}\small [hep-th]}].

\bibitem{Jafferis:2010un}
D. L. Jafferis,
``The Exact Superconformal R-Symmetry Extremizes Z'',
\jhep{1205}{2012}{159} [\arXivid{1012.3210} {\color{cyan}\small [hep-th]}].

\bibitem{Hama:2010av}
N. Hama, K. Hosomichi and S. Lee,
``Notes on SUSY Gauge Theories on Three-Sphere'',
\jhep{1103}{2011}{127} [\arXivid{1012.3512} {\color{cyan}\small [hep-th]}].

\bibitem{Kallen:2012cs}
J. Kallen and M. Zabzine,
``Twisted Supersymmetric 5D Yang-Mills Theory and Contact Geometry'',
\jhep{1205}{2012}{125} [\arXivid{1202.1956} {\color{cyan}\small [hep-th]}].

\bibitem{Jafferis:2012iv}
D. L. Jafferis and S. S. Pufu,
``Exact Results for Five-dimensional Superconformal Field Theories with Gravity Duals'',
2012, \arXivid{1207.4359} {\color{cyan}\small [hep-th]}.

\bibitem{Jafferis:2011zi}
D. L. Jafferis, I. R. Klebanov, S. S. Pufu and B. R. Safdi,
``Towards the F-Theorem: N=2 Field Theories on the Three-Sphere'',
\jhep{1106}{2011}{102} [\arXivid{1103.1181} {\color{cyan}\small [hep-th]}].

\bibitem{Maldacena:1997re}
 J. M. Maldacena,
 ``The Large N Limit of Superconformal Field Theories and Supergravity'',
\emph{Adv. Theor. Math. Phys.} {\bf 2} (1998) 231 [\arXivid{hep-th/9711200}].
 
\bibitem{Gubser:1998bc}
S. S. Gubser, I. R. Klebanov and A. M. Polyakov,
``Gauge Theory Correlators from Noncritical String Theory'',
\pl{B}{428}{1998}{105} [\arXivid{hep-th/9802109}].

\bibitem{Witten:1998qj}
E. Witten,
``Anti-de Sitter Space and Holography'',
\emph{Adv. Theor. Math. Phys.} {\bf 2} (1998) 253 [\arXivid{hep-th/9802150}].

\bibitem{Drukker:2010nc}
N. Drukker, M. Marino and P. Putrov,
``From Weak to Strong Coupling in ABJM Theory'',
\cmp{306}{2011}{511} [\arXivid{1007.3837} {\color{cyan}\small [hep-th]}].

\bibitem{Aharony:2008ug}
O. Aharony, O. Bergman, D. L. Jafferis and J. Maldacena,
``N=6 Superconformal Chern-Simons-matter Theories, M2-branes and their Gravity Duals'',
\jhep{0810}{2008}{091} [\arXivid{0806.1218} {\color{cyan}\small [hep-th]}].

\bibitem{Klebanov:1996un}
I. R. Klebanov and A. A. Tseytlin,
``Entropy of Near Extremal Black p-branes'',
\npb{475}{1996}{164} [\arXivid{hep-th/9604089}].

\bibitem{Herzog:2010hf}
C. P. Herzog, I. R. Klebanov, S. S. Pufu and T. Tesileanu,
``Multi-Matrix Models and Tri-Sasaki Einstein Spaces'',
\pr{D}{83}{2011}{046001} [\arXivid{1011.5487} {\color{cyan}\small [hep-th]}].

\bibitem{Martelli:2011qj}
D. Martelli and J. Sparks,
``The Large N Limit of Quiver Matrix Models and Sasaki-Einstein Manifolds'',
\pr{D}{84}{2011}{046008} [\arXivid{1102.5289} {\color{cyan}\small [hep-th]}].

\bibitem{Cheon:2011vi}
S. Cheon, H. Kim and N. Kim,
``Calculating the Partition Function of N=2 Gauge Theories on $S^3$ and AdS/CFT Correspondence'',
\jhep{1105}{2011}{134} [\arXivid{1102.5565} {\color{cyan}\small [hep-th]}].

\bibitem{Jafferis:2008qz}
D. L. Jafferis and A. Tomasiello,
``A Simple Class of N=3 Gauge/Gravity Duals'',
\jhep{0810}{2008}{101} [\arXivid{0808.0864} {\color{cyan}\small [hep-th]}].

\bibitem{Gulotta:2011si}
D. R. Gulotta, C. P. Herzog and S. S. Pufu,
``From Necklace Quivers to the F-theorem, Operator Counting, and T(U(N))'',
\jhep{1112}{2011}{077} [\arXivid{1105.2817} {\color{cyan}\small [hep-th]}].

\bibitem{Yee:2006ba}
H.-U. Yee,
``AdS/CFT with Tri-Sasakian Manifolds'',
\npb{774}{2007}{232} [\arXivid{hep-th/0612002}].

\bibitem{Assel:2012cj}
B. Assel, C. Bachas, J. Estes and J. Gomis,
``IIB Duals of D=3 N=4 Circular Quivers'',
2012, \arXivid{1210.2590} {\color{cyan}\small [hep-th]}.
  
\bibitem{Gulotta:2011vp}
D. R. Gulotta, J. P. Ang and C. P. Herzog,
``Matrix Models for Supersymmetric Chern-Simons Theories with an ADE Classification'',
\jhep{1201}{2012}{132} [\arXivid{1111.1744} {\color{cyan}\small [hep-th]}].

\bibitem{Willett:2011gp}
B. Willett and I. Yaakov,
``N=2 Dualities and Z Extremization in Three Dimensions'',
2011, \arXivid{1104.0487} {\color{cyan}\small [hep-th]}.

\bibitem{Benini:2011mf}
F. Benini, C. Closset and S. Cremonesi,
``Comments on 3d Seiberg-like Dualities'',
\jhep{1110}{2011}{075} [\arXivid{1108.5373} {\color{cyan}\small [hep-th]}].

\bibitem{Aharony:1997gp}
O. Aharony,
``IR Duality in d=3 N=2 Supersymmetric USp(2N(c)) and U(N(c)) Gauge Theories'',
\pl{B}{404}{1972}{71} [\arXivid{hep-th/9703215}].

\bibitem{Giveon:2008zn}
A. Giveon and D. Kutasov,
``Seiberg Duality in Chern-Simons Theory'',
\npb{812}{2009}{1} [\arXivid{0808.0360} {\color{cyan}\small [hep-th]}].

\bibitem{SignedGraphs1}
F. Harary,
``On the Notion of Balance of a Signed Graph'',
\emph{Michigan Math. J.} \href{http://projecteuclid.org/euclid.mmj/1028989917}{{\bf 2} (1953-54) 143}.

\bibitem{SignedGraphs2}
Thomas Zaslavsky,
``The Geometry of Root Systems and Signed Graphs'',
\emph{The American Math. Monthly} \href{http://www.jstor.org/stable/2321133}{{\bf 88} (1981) 88}.

\bibitem{SignedGraphs3}
Thomas Zaslavsky,
``Signed Graphs'',
\emph{Discrete Applied Math.} \href{http://www.sciencedirect.com/science/article/pii/0166218X82900336}{{\bf 4} (1982) 47}.

\bibitem{SignedGraphs4}
Seth Chaiken,
``A Combinatorial Proof of the All Minors Matrix Tree Theorem'',
\emph{SIAM. J. on Algebraic and Discrete Methods}, \href{http://dx.doi.org/10.1137/0603033}{{\bf 3} (1982) 319}.
 
\bibitem{Klebanov:2011gs}
I. R. Klebanov, S. S. Pufu and B. R. Safdi,
``F-Theorem Without Supersymmetry'',
\jhep{1110}{2011}{038} [\arXivid{1105.4598} {\color{cyan}\small [hep-th]}].
  
\bibitem{Amariti:2011da}
A. Amariti and M. Siani,
``Z-extremization and F-theorem in Chern-Simons Matter Theories'',
\jhep{1110}{2011}{016} [\arXivid{1105.0933} {\color{cyan}\small [hep-th]}].

\bibitem{Klebanov:2011td}
I. R. Klebanov, S. S. Pufu, S. Sachdev and B. R. Safdi,
``Entanglement Entropy of 3-d Conformal Gauge Theories with Many Flavors'',
\jhep{1205}{2012}{036} [\arXivid{1112.5342} {\color{cyan}\small [hep-th]}].

\bibitem{Casini:2011kv}
H. Casini, M. Huerta and R. C. Myers,
``Towards a Derivation of Holographic Entanglement Entropy'',
\jhep{1105}{2011}{036} [\arXivid{1102.0440} {\color{cyan}\small [hep-th]}].

\bibitem{Casini:2012ei}
H. Casini and M. Huerta,
``On the RG Running of the Entanglement Entropy of a Circle'',
\pr{D}{85}{2012}{125016} [\arXivid{1202.5650} {\color{cyan}\small [hep-th]}].

\bibitem{Gulotta:2012yd}
D. R. Gulotta, C. P. Herzog and T. Nishioka,
``The ABCDEF's of Matrix Models for Supersymmetric Chern-Simons Theories'',
\jhep{1204}{2012}{138} [\arXivid{1201.6360} {\color{cyan}\small [hep-th]}].
}
\end{document}